\documentclass{article}
\usepackage{fullpage}
\usepackage{amsfonts,amssymb}
\usepackage{amsthm}
\usepackage{tikz}
\usetikzlibrary{calc}
\usetikzlibrary{arrows}
\usepackage{tikz-3dplot}
\usepackage{color} 
\usepackage[fleqn]{amsmath}
\usepackage{amsthm}
\usepackage{amssymb}
\usepackage{amsfonts}
\usepackage{bbm}
\usepackage{epsfig}
\usepackage{times}
\usepackage{hyperref}
\usepackage{bm}
\usepackage{float} 
\usepackage{appendix} 
\usepackage{lscape}
\usepackage{cite}
\usepackage{mathrsfs}
\usepackage{appendix}
\usepackage{setspace}
\usepackage{mathtools}
\usepackage{authblk} 

\theoremstyle{definition}
\newtheorem{definition}{Definition}[section]

\begin{document}
 
	 \title{Is There Causation in Fundamental Physics? New Insights from Process Matrices and Quantum Causal Modelling } 
	 
 \author{Emily Adlam  \thanks{The Rotman Institute of Philosophy, 1151 Richmond Street, London N6A5B7 \texttt{eadlam90@gmail.com} }}

	 \date{\today}

	 \maketitle

Bertrand Russell famously argued that causation plays no role in science: it is `\emph{a relic of a bygone age, surviving, like the monarchy, only because it is erroneously supposed to do no harm.}’\cite{10.2307/4543833}  Cartwright\cite{10.2307/2215337} and later writers moderated this conclusion somewhat, and it is now largely accepted that in a macroscopic setting causal concepts are an important part of the assessments we make about possible strategies for action. But the view that causation in the usual sense of the term is not present in fundamental physics, or at least that not all fundamental physical processes are causal, remains prevalent\cite{Woodward2007-WOOCWA, Field2003-FIECIA} - for example, Norton writes that `\emph{(causes and causal principles) are heuristically useful notions, licensed by our best sciences, but we should not mistake them for the fundamental principles of nature}'\cite{Norton2003-NORCAF}. Furthermore, many influential philosophical analyses of causation posit that causation arises only at a macroscopic level,  as a result of the thermodynamic gradient\cite{Lewis1979-LEWCDA,10.2307/20012433}, interventions\cite{pearl2009causality,woodward2005making}, the perspectives of agents\cite{Price2005-PRICP}, or some such feature of reality which plays no role in fundamental physics.  

In light of this widespread orthodoxy, it may seem surprising that in recent years a significant literature around causation has sprung up within quantum foundations.  Thus it is important to understand the nature of the recent renaissance of causal talk in fundamental physics: do these research programmes offer a counterexample to the claim that causal notions do not appear in fundamental physics, or are they simply using the word `cause' in a non-standard way? In this article, we address this question in the context of two research programmes in quantum foundations: the process matrix formalism, and the causal modelling approach. 

 Woodward\cite{10.1086/678313} distinguished between several different philosophical projects concerning our understanding of causation - a descriptive project, a metaphysical project, and  a functional project. This article spans all three of these projects. We begin with the  descriptive project, aiming to understand what physicists working in the process matrix formalism and the causal modelling approach mean by their use of causal terminology; thus we review a debate between proponents of the two approaches over the significance of an experiment called `the quantum SWITCH,' which brings into sharp focus the differing conceptions of causation employed in these two frameworks. We show that   the  process matrix  programme has correctly identified a  notion of \emph{causal order} which plays an important role in fundamental physics, and argue that this notion is weaker than  the common-sense conception of causation because it does not demand any kind of asymmetry. 

We then take up the metaphysical project, arguing that  causal order plays an important role in grounding more familiar causal phenomena; thus we conclude that Russell was correct in his observation that there is no  causation in the usual sense in fundamental physics, but nonetheless fundamental physics does exhibit a phenomenon which is closely related to causation.  We  apply these conclusions to assess the prospects of the causal modelling approach within quantum foundations: since no-signalling quantum correlations cannot exhibit causal order, we argue that certain quantum phenomena should not be analysed using classical causal models. This resolves an open question about how to interpret fine-tuning in classical causal models of no-signalling correlations.  

Finally we move to the functional project, where we observe that a quantum generalization of causal modelling can play a similar functional role to standard causal reasoning. Thus we conclude that quantum causal  models have a legitimate claim to be regarded as at least quasi-causal despite the absence of causal order in the phenomena they describe; but we emphasize that this functional characterisation does not entail that quantum causal models offer novel explanations of quantum processes.

 \section{Background: Process Matrix Formalism\label{PM}} 
 
\begin{figure}
	\centering
	\begin{tikzpicture}[scale=0.7]

	\node[below, black] at (-1,1.25) { T}; 
		\coordinate (a) at (-2,0);
	\coordinate (b) at (-2,2);
	\coordinate (c) at (0,2);
	\coordinate (d) at (0,0);

	\draw[black] (a) -- (b);
	\draw[black] (b) -- (c);
	\draw[black] (c) -- (d);
	\draw[black] (d) -- (a);
	
	\node[below, black] at (-1,4.5) { C};

	\draw[ultra thick, ->] (0,1) arc (-60:240:2);

	\end{tikzpicture}	
	\caption{Schematic diagram of a channel  around a closed causal loop with one laboratory}
	\label{signalling}
\end{figure}
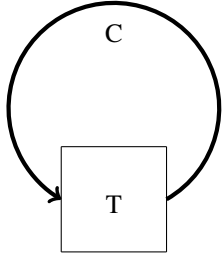

 To understand the purpose of the process matrix formalism, let us begin by considering the classical case.  Using purely classical physics we can (in principle) construct  processes which lack a unique causal order by putting our laboratories on a closed timelike curve. For example, suppose we have a single lab on a closed timelike curve, with its output connected to its input via a channel $C$, as shown in figure \ref{signalling}. An obvious problem with this kind of setup is that it can lead to apparent logical contradictions. For example, suppose that $C$ is the identity channel whilst $T$ is a bit flip operation; then necessarily the output of the lab equals both $0$ and $1$ at once, which most people would consider to be both logically and physically impossible.  The problem can be avoided if we say that the agents in the lab are simply not able to perform a bit flip operation, but this is unappealing since we typically think that classical agents should be able to perform any classically allowed local operation within an isolated lab.

The classical process function formalism\cite{reversible} has been formulated to address this kind of problem. First, we consider a set of agents in separate laboratories who are free to perform any operation which is permitted by classical physics; then we define a `process function' to describe the classical dynamics between the labs, without making any assumptions about the causal ordering of the labs. The set of valid process functions is defined  in such a way as to include all possible dynamics which will never give rise to contradictions no matter which operations the individual agents choose to perform.  For example, the identity channel from a lab to itself not one of the allowed dynamics, because it can lead to the kind of contradiction we have just described. In fact, there are no valid process functions on a CTC involving one or two labs, apart from trivial dynamics which output a constant or random value. However it can be shown that for three or more labs on a CTC there are nontrivial process functions, thus demonstrating that there are possible classical dynamics on a CTC which allow agents complete local freedom whilst avoiding contradictions  (here the term `possible' should be understood in the sense of `logically possible' - it is not yet known whether dynamics such as these can be implemented in our actual world). 

The process matrix formalism\cite{Oreshkov2,Oreshkov,Ara_jo_2017,rubinorozema,Goswami_2018} is simply the quantum version of this formalism. We define a `process' to be composed of a set of  agents in separate laboratories who are free to perform any local operation which is permitted by quantum mechanics, together with a  `process matrix' which describes the quantum dynamics between the labs, without making any assumptions about the causal ordering of the labs. The process matrix operates as a  higher-order map from a set of quantum operations to another operation - i.e. it maps the quantum operations performed by the agents in the laboratories to  an operation representing the net effect of performing all of those operations in the context of the particular process. The set of valid process matrices is defined  in such a way as to include all possible dynamics which will never give rise to contradictions no matter which operations the individual agents choose to perform.  Again we find that there are possible dynamics which don't have a well-defined causal order but which allow agents to perform all operations allowed by quantum physics; but in the quantum case we have more possibilities for non-causal structures - we could still have CTCs,  but we could also have superpositions of different causal structures, which are often characterised as `indefinite causal order.'

\begin{figure}
	\centering
\begin{tikzpicture} 
\draw (0,0) circle (30pt);
\node[below, black] at (0,0.5) { causally}; 
\node[below, black] at (0,0) { ordered}; 
\draw (0,0) circle (80pt);
\node[below, black] at (-1.9,0.5) { causally}; 
\node[below, black] at (-1.9,0) { separable}; 
\draw (0,0) circle (130pt);
\node[below, black] at (-3.75,0.25) { causal};  

\node[below, black] at (-7,0.5) { non-causal   processes};  
\node[below, black] at (-7,0) { (not yet observed in reality) };  

\filldraw (3,0.75) circle (2pt);
\node[below, black, ] at (3.5,0.75) {\scriptsize Quantum };  
\node[below, black] at (3.5,0.5) {\scriptsize SWITCH };  

\end{tikzpicture}
	\caption{Schematic diagram of  types of process in the process matrix formalism}
	\label{cause}
\end{figure}
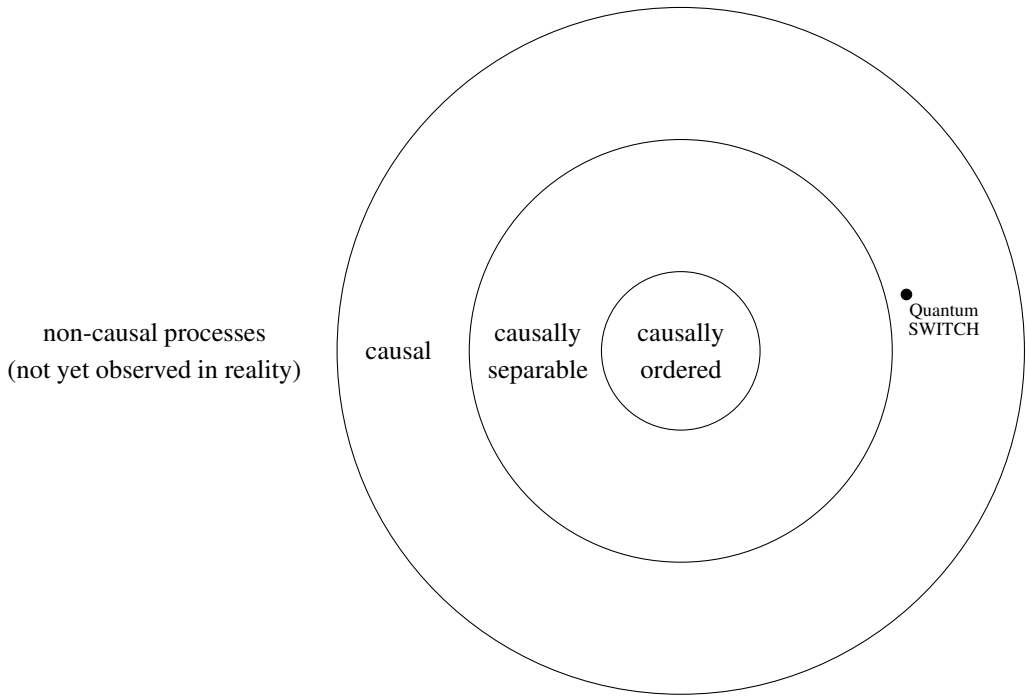

Process matrices are classified in terms of the degree of causal structure that they exhibit, as shown in figure \ref{cause}.  First, we have the process matrices which are   \emph{causally ordered}, i.e. they are compatible with the existence of a strict partial order  such that   the probabilities for the outcomes in some laboratory depend  only on the settings  in that laboratory or in  experiments which precede it in the partial order. Note that here and throughout this article, `depend' refers to statistical dependence, i.e. it encodes the possibility of signalling - the kinds of relationships that may occur between no-signalling laboratories, as in the Bell correlations, do not count for the purpose of these definitions. Second, we have process matrices which are \emph{causally separable}, i.e. they can be written as a convex decomposition of causally ordered process matrices. Third, we have process matrices which are neither causally ordered nor causally separable, but are still \emph{causal} in the sense that we can write down a well-defined probability distribution over possible strict partial orders such that in each possible case, the choice of setting in a local experiment does not affect the occurrence of events which are earlier in the order, nor the strict partial order on the set of the events and the experiment in question\cite{articleoreshkov}.\footnote{ More precisely, according to ref \cite{articleoreshkov} a process matrix is causal iff there exists a probability distribution over possible strict partial orders on the set of local experiments and outcomes of those experiments such that for every local experiment $A$, every subset $X$ of the other local experiments, and every choice of strict partial order, the joint probability distribution over the outcomes for the experiments $X$ and the possible strict partial orders over $A, X$ (obtained by summing over only those orders for the full process such that    $A$ does not precede any experiments in $X$) are independent of the settings in the experiment $A$.} Finally, we have the process matrices which are \emph{non-causal}, meaning that they are not compatible with a distribution over strict partial orders which obeys these requirements. We can define `causal inequalities,' somewhat analogous to the Bell inequalities which can only be violated by non-causal processes\cite{Branciard_2015}.  Although a number of  non-causal process matrices have been formulated and studied, all processes currently known to be realised in nature (without post-selection) are causal.

Ref \cite{renato} gives greater insight into the reasons why all processes that we have so far been able to implement are causal: in fact, any process made up of localised events which is implemented in a fixed acyclic background spacetime in a way which satisfies relativistic causality (i.e. no superluminal signalling) can be  rewritten using a `fine-graining' process which splits individual laboratories into several distinct laboratories in order to produce  a fixed  acyclic causal structure. This makes sense, since  after all the acyclic  background spacetime necessarily imposes a global strict partial order on any processes performed in that spacetime, if those processes are required to respect relativistic causality. In particular, even if the process involves quantum systems in a superposition of different causal histories, relativistic causality entails that `\emph{an agent has the potential to intervene at any of the spacetime locations to verify the probability of detecting the particle there ... (and) if agents choose to perform such interventions, this will not enable them to signal outside the future of the spacetime,}' and it is this requirement which ensures that these processes can always be rewritten with a fixed causal order.   Ref  \cite{renato} therefore argues that no experiment in a well-defined spacetime can demonstrate `indefinite causal order' in a strong sense - in their view, the only way to exhibit  genuine `indefinite causal order' would be to do some kind of quantum gravity experiment where the spacetime background is no longer fixed. However, this still leaves open questions about how to interpret processes performed with a  fixed background which   at least superficially seem to involve superpositions of causal orders; we will return to this question in section \ref{SWITCH}.

\subsubsection{Causal terminology \label{probs}}

 Often in the process matrix framework the term `causal' seems to refer simply to the transmission of information between systems, so for example the simplest kind of `causally ordered' processes are just one-way signalling processes - in effect $A$ is regarded as being a cause of $B$ if and only if $p(B | A) \neq p(B)$\footnote{Some proponents of the process matrix formalism seem inclined to identify causation with signalling\cite{articleoreshkov}, but others take a more nuanced approach - for example,  ref \cite{renato} acknowledges that signalling is not a necessary condition for the presence of a causal relation, as one can imagine classical cases where two causal influences exactly cancel one another out, so there is no observable signalling effect even though there are cause-effect relations. Thus ref \cite{renato} merely uses signalling as a sign that causation is present, rather than a definition of causation.}. This  information-theoretic way of thinking about causation is essentially a version of Lewis' simplest probabilistic counterfactual theory of causation\cite{Lewis1986-LEWPTC}, where we say that $A$ is a cause of $B$ if and only if, if $A$ had not occurred, the chance of $B$ occurring would be less than its actual chance.  In the process framework we don't particularly care about whether $A$ makes $B$ more or less probable as long as it changes the probability of $B$, but it is clear that the approaches are similarly motivated and will have similar advantages and disadvantages. In particular, it is well-recognised\cite{sep-causation-counterfactual} that the simplest  counterfactual account  is not asymmetric and hence must rely on contingent external considerations (e.g. thermodynamics) if it is to recover the asymmetry of causation,    and evidently the process matrix formulation faces a similar problem, since a relation defined in terms of the possibility of signalling between two variables is evidently not inherently asymmetric. 

Approaches to causality aiming to reproduce its asymmetry often do so by appealing directly to agency or agents  - for example, interventionists  argue that a variable A causes a variable B if and only if B would change if A were manipulated by an agent in an appropriate way\cite{pearl2009causality,woodward2005making}. The interventionist approach may seem like a good fit for the process matrix formalism, since the operations performed by the agents in the laboratories can be regarded as interventions. However, problems arise from the fact that the  literature on process matrices  has a tendency to equivocate around the nature of the `laboratories' in which the quantum operations are performed. On the one hand, the experiments are described in terms of agents - that is to say, they are usually said to be performed by entities  named `Alice' and `Bob,' and readers could be forgiven for assuming that Alice and Bob are supposed to be human beings. But in fact, in practical implementations of indefinite causal order, such as the quantum SWITCH (see section \ref{SWITCH}), the roles of `Alice' and `Bob' are played by quantum gates or similar quantum instruments - and indeed, certain interpretations of quantum mechanics, like the spontaneous collapse approach\cite{sep-qm-collapse} \footnote{We use the term `interpretation' in the popular sense here, as referring to any possible solution to the measurement problem, e.g. Everett, de Broglie-Bohm, GRW collapse and so on.  It might be argued that not all of these approaches are really `interpretations' - de Broglie Bohm and GRW, for example, add something to standard quantum mechanics and thus should perhaps be referred to as `alternative hypotheses.' However we use the term `interpretation' for consistency with the existing literature.}   have the consequence that it is effectively impossible for an agent to be in a superposition, which would entail that a `quantum SWITCH' and other similar processes could \emph{never} be implemented with a real agent. And this distinction is not trivial, because if we want to interpret the `causal structures' involved here in the interventionist sense,  then we must insist that the operations performed by Alice and Bob are genuinely `free' interventions; but quantum gates evidently do not exercise freedom of choice, so it is less clear that the causal interpretation is still valid when these experiments are implemented using quantum gates.   

Furthermore, in order to employ an interventionist account of causation we will need to be able to assign probabilities to measurement outcomes conditional on the actions taken by various possible agents; and although it is straightforward to write down such probabilities, it is less clear what the probabilities actually mean,  because if we to keep the agents in coherent superposition then we will not be able to invoke a collapse interpretation to make sense of them. Perhaps the most well-developed extant approach to making sense of quantum probabilities in a no-collapse setting is the Deutsch-Wallace decision-theoretic analysis\cite{https://doi.org/10.48550/arxiv.0906.2718,Wallacebook} developed in the context of the Everett interpretation; but this approach may not be applicable in the context of  experiments probing indefinite causal order, because it's designed to be applied in the kind of regime where different branches of the wavefunction have effectively decohered and thus interference between branches is impossible, and this is not the case in experiments like the quantum SWITCH because the point of these experiments is that the branches are kept in coherent superposition and ultimately recombined (otherwise we could not verify that the superposition of causal structure has taken place). For example, one important axiom in the decision-theoretic analysis  is `\emph{Branching Indifference: An agent doesn’t care about branching per se: if a certain measurement leaves his future selves in N different macrostates but doesn’t change any of their rewards, he is indifferent as to whether or not the measurement is performed}'\cite{Wallacebook} - this axiom seems very reasonable  if branches can't interfere, but not so much in the case where agents remain in coherent superposition, because in that case branching brings with it the possibility of phenomena like interference and recombination which could  lead to the agents' memories being altered or wiped, and these certainly seem like things they might care about! So it is unclear that we can define a sensible notion of probability for agents involved in experiments like the quantum SWITCH.

In virtue of these concerns,  it should not be assumed that terms like `causally ordered' and `causal'  in the process matrix framework line up exactly with  the common-sense usage or with any standard philosophical analysis of causation. Thus for the moment terms like `causally ordered' and `causal'  should be understood in a purely formal way. One of our aims in this article is to determine the connections between these technical notions and the more general concept of causality, so we will eventually have something more specific to say about the sense in which `causal order' really pertains to `causation' in the ordinary sense.

 \section{Background: Causal Modelling \label{causalmodel}}
 
 The classical causal modelling framework, whose developers and proponents include  Pearl\cite{pearl2009causality} and Spirtes, Glymour, and Scheines\cite{spirtes2000causation}, is a methodology for making inferences about causal relationships from statistical data. A causal model for a set of variables is a directed graph whose edges represent causal relationships between the variables $\{X_i\}$, together with a set of conditional probabilities  $\{  p(X_i | Pa(X_i)) \}$ giving the probability distribution for the variable   $X_i$ conditioned only on the set $Pa(A_i)$ of its parents in the directed graph (where X is a parent of Y iff there is an arrow from   X to Y in the graph). The probabilities are supposed to  obey a `Markov condition,' which states that the joint probability distribution over all the variables   can be written as the product of each of the individual probability distributions  $\{  p(X_i | Pa(X_i)) \}$. It is usually assumed that the directed graph must be acyclic, although in some circumstances this assumption is relaxed\cite{https://doi.org/10.48550/arxiv.1309.6836,Neal_2000}.  The causal modelling framework is closely linked to the interventionist account of causation\cite{pearl2009causality,woodward2005making}, as causal models can be regarded as encoding  the results of possible interventions.

The causal modelling approach was originally intended for the study of macroscopic causal relationships, such as the consequences of medical interventions. In this setting it is uncontroversial that the  variables stand in causal relationships, since the macroscopic world exhibits many causal features which we must take seriously if we wish to characterise  `\emph{an ‘objective’ distinction between effective and ineffective strategies'}\cite{10.2307/2215337}. However, in recent years there has been increasing interest   in applying the tools of causal modelling to the microscopic world, and in particular to  quantum mechanics. One might question whether techniques designed to study macroscopic causality can be sensibly translated to this  context, given the aforementioned doubts  about whether causation features in fundamental physics, so one of our aims in this article will be to reach a verdict on this matter. 
 
 Causal modelling has been used to analyse various quantum phenomena. For example,  one significant result in this framework is due to Wood and Spekkens\cite{SpekkensWood}, who showed that  any classical causal model which reproduces the Bell correlations must be `fine-tuned' in the sense that the underlying causal model postulates causal connections between events which appear causally unconnected at the level of the empirical statistics. To see how this works, note that the  proof of the Bell inequality can be regarded as telling us that any causal model for correlations violating the Bell inequality must postulate a causal connection between the choice of measurement on one particle and the outcome of the measurement on the other particle, but the quantum mechanical no-signalling theorem ensures that at the statistical level there will be no dependence of the outcome on the measurement choice, so if we wish to represent these statistics by a causal model we must carefully `fine-tune' the parameters of the model to ensure that the underlying causal influences exactly cancel out so as to be invisible at the level of the empirical statistics.

 There also exists a generalisation of Pearl's schema to quantum mechanics which replaces classical causes with quantum ones. Various versions of this approach exist\cite{Chaves,Miklin_2017,Pienaar_2017}, but here we will follow the approach of ref \cite{https://doi.org/10.48550/arxiv.1906.10726} which takes `\emph{causal relations to be relations of influence between quantum nodes mediated by unitary evolution.}'  In this approach, a quantum causal model is defined on a set of quantum nodes - each node  is a pair of Hilbert spaces, one associated with   an incoming system, and one associated with an outgoing system, and the idea is that an agent could potentially perform an operation  $O_i$ at a node $A_i$, thus implementing a transformation on the incoming system to produce the outgoing system.  Thus a quantum causal model constitutes a directed acyclic graph on a set of quantum nodes, together with a set of quantum channels $\{ \rho_{A_i | Pa(A_i)} \}$ such that the incoming system for node $A_i$ is given by the output of the channel $\rho_{A_i | Pa(A_i)}$ acting on the set $Pa(A_i)$ of its parents in the DAG, and every channel  $\rho_{A_i | Pa(A_i)}$ commutes with every other channel $ \rho_{A_j | Pa(A_j)}$. The process as a whole can be described by a `process operator' which maps the set of operations $O_i$ implemented by observers at nodes to an operation which represents the net effect of performing all of these operations in the context of the particular process; usually it is assumed that the channels should obey a `quantum Markov condition,' which requires that the process operator   can be written as the product of each of the individual channels $\rho_{A_i | Pa(A_i)}$. It can be seen that this approach has much in common with the process matrix framework - the `nodes' play the role of laboratories where agents freely choose their operations, and the `process operator' plays a similar role to the process matrix. However, the emphasis here is on defining a specific causal structure rather than allowing indefinite causal orders. 
 
 If we assume that a set of operationally defined statistics is produced by a quantum causal structure taking this specific form, we can then use statistical tools to make inferences about the nature of the underlying quantum structure - for example, given a process operator we can use the algorithm set out in ref \cite{https://doi.org/10.48550/arxiv.1906.10726} to find out if the operator obeys the Markov condition for some particular DAG on the quantum nodes, thus determining whether that DAG represents a candidate quantum causal structure for the process.  Evidently there are a number of practical uses for this formalism - for example, given an experimental procedure where we are not sure our apparatus is doing exactly what we expected, we can apply quantum causal inference to determine if the observed statistics are consistent with our desired underlying quantum structure. However, it has also been suggested that there may   be foundational applications of this formalism - it serves to `\emph{establish an account of causality in quantum theory’s own terms, without assuming a separate realm of classical systems or measurement outcomes. }'\cite{https://doi.org/10.48550/arxiv.1906.10726}  One of our aims in this article will be to evaluate these claims.

 \section{The Descriptive Project: The Quantum SWITCH  \label{SWITCH}} 
 
Let us now begin with the descriptive project, seeking to understand what exactly physicists working in the process matrix research programme mean by their use of causal terminology. To do this, we will examine the discussion around an experiment known as the quantum SWITCH, which is a causally non-separable but still causal process originally proposed as an example of indefinite causal structure  in the process matrix formalism\cite{Chiribella_2013, Goswami_2018}.   In this process, we use a  control qubit to determine the order in which two gates $A$, $B$ are applied to some other qubit: if the control qubit is in the state $| 0 \rangle$ the gate $A$ is applied first and the gate $B$ follows, whereas if the control qubit is in the state $| 1 \rangle$ the gate $B$ is applied first and the gate $A$ follows. Evidently then if the control qubit is prepared in a superposition of the state $| 0 \rangle$ and the state $| 1 \rangle$ we will get a superposition of two orderings - in one branch the gates will be applied in the order $A, B$ and in the other the gates will be applied in the order $B, A$. This scenario is interesting particularly because it can   be  implemented with current technology, thus it has already been demonstrated in the lab\cite{Goswami_2018}.
 
The conceptual significance of the quantum SWITCH has been questioned by physicists working on the causal modelling approach. In particular, in ref \cite{MacLean_2017} MacLean  et al make two objections to the claim that the quantum SWITCH represents a genuine superposition of two different causal orders. The first is that the two branches do not really contain copies of the same events in different orders; the events occur at different times in each of the branches, so really we should think of them as completely distinct events. Gu\'{e}rin and Brukner\cite{Gu_rin_2018} respond to this by  invoking diffeomorphism invariance to argue that events cannot be identified across branches by their spacetime location, since that is not a diffeomorphism-invariant fact - instead they must be identified  by the intrinsic properties of the event, which are the same in each branch of the quantum SWITCH. Now, this argument might succeed if it these events were occurring in complete isolation so that there were no  way to identify them other than either spacetime location or the intrinsic properties of the event. But in this case the events are not isolated: they are controlled by an external system and thus they stand in well-defined relations to that external system, meaning that there are many ways they can be identified by means of their relations to other external events. Diffeomorphism invariance is thus not really a relevant consideration in this kind of experiment. 

However, although the diffeomorphism argument does not succeed, there may be other reasons to conclude that we should think of the two branches as containing `the same events.' To get clearer on this question,  it is helpful   to consider in more detail what is actually achieved by the quantum SWITCH. On the one hand,  the results of refs  \cite{Paunkovi__2020, renato} underline that  the quantum SWITCH does not exhibit the strongest and most interesting kind of `indefinite causal order' that is possible within the process matrix framework\footnote{Ref \cite{Paunkovi__2020} demonstrates that the  quantum SWITCH is not equivalent to the gravitational SWITCH, which implements a similar circuit but uses superpositions of different spacetime geometries to determine the two different orders; this indicates that the ordinary quantum SWITCH does not exhibit indefinite causal order in the same unambiguous way as the gravitational SWITCH. Meanwhile, the results of \cite{renato} show that at the purely operational level it would be possible to come up with a `fine-grained' version of the quantum SWITCH process which can be assigned a well-defined causal order, so operationally speaking the SWITCH is not achieving anything that could not be achieved with a fixed causal order.}. On the other hand,  if Maclean et al were right that in fact completely different events occur in each branch, then one would naturally expect the quantum SWITCH to be equivalent to a superposition in which the qubit goes down two different paths, one containing the gate $A$ and the gate $B$, and another containing a completely distinct copy of the gate $B$ and then a completely distinct copy of the gate $A$.  However,  as shown in ref \cite{doi:10.1098/rspa.2018.0903}, the quantum SWITCH is not equivalent to this superposition - rather it is equivalent to a superposition in which the two paths are correlated. That is to say, the quantum SWITCH produces the same results as a scenario where we have one  spacetime region containing a system $S_A$ which implements gate $A$ and another spacetime region containing a system $S_B$ which implements gate $B$, and in one branch the qubit first visits the $A$-region and then the $B$-region, while in the other branch the qubit first visits the $B$-region and then the $A$-region. In this case it is easy to see that the reason the paths become correlated is precisely because the copy of gate A that the qubit visits  is really \emph{the same gate} in each branch: in the branch where the qubit visits the $A$-region first, it interacts with $S_A$ and undergoes gate $A$, and meanwhile the state of $S_A$ is presumably altered, but in the other branch nothing happens during that period of time and instead the qubit visits the $A$ region later, finding $S_A$ still in its original quantum state so that gate $A$ is implemented.  So this equivalence seems to uphold the conclusion of Gu\'{e}rin and Brukner that the quantum SWITCH is in some sense  a superposition of different causal orders.

 The second objection made by MacLean et al is that even if we agree that the same events occur in the two branches, nonetheless we don't really have indefinite causal order, because the causal structure is the same in both branches. That is to say, in both branches the causal structure is `$C$ causes both $A$ and $B$,' where $C$ is the control qubit. And indeed, this does seem to be a correct description of the  experiment, so this objection seems like it could be a serious problem for the indefinite causal order programme - and even more so because Costa has recently shown that under quite broad conditions it is not possible to have a valid process in which we have a pure superposition of causal orders without any external control\cite{Costa_2022}, so the same objection will hold even in more complex examples of  indefinite causal order. 
 
 However it does not seem that the proposers of the quantum SWITCH ever meant to claim that this experiment represents a superposition of different causal  \emph{structures} in the sense in which MacLean et al use this term.  Rather they argued that it demonstrates a superposition of different  \emph{causal orders}\cite{Chiribella_2013}. In fact there is some confusion here because the language used by researchers in this programme equivocates between `causal order,' `causal structure' and `causal relations.' In some cases it seems to be implied that all three are the same: for example ref \cite{10.3389/fphy.2020.525333} seems to be making an identification between `causal order' and `causal relations' in their reference to `\emph{realization of indefinite causal order (ICO), a theoretical possibility that even causal relations between physical events can be subjected to quantum superposition.}' But on further  examination, it is clear that these things are not identical. The term `causal relations' is naturally understood as referring to `cause-effect relations' of the kind that we are familiar with from our macroscopic experiences, and the term `causal structure,' usually seems to refer to causal structures as analysed in the causal modelling framework, i.e. structures made up out of these `cause-effect relations.' \emph{Causal order}, on the other hand, does not invariably refer to cause-effect relations. The quantum SWITCH is a good example: for after all, even if we ignore the control qubit it is clearly not the case that in one branch $A$ \emph{causes} $B$ and in the other $B$ \emph{causes} $A$ - the experiment is not set up in such a way that the application of the first gate is `the cause' of the application of the second gate. But nonetheless the order in which these events occur is meaningful,  otherwise the two branches would be identical and we would not have a superposition at all. And moreover, we have seen that there are indeed good reasons to think that it really is the \emph{order} of events which differs between these branches in significant ways.   So it seems reasonable to say that these events stand in a physically meaningful \emph{causal order} even though they do not bear any cause-effect relation to one another.

\subsection{Causal Order} 

The observations of the previous section provide good reason to think that `causal order' plays an important role in the process matrix formalism. But what exactly is causal order, and how does it differ from the more familiar notion of cause-effect relations?  Though it may be tempting to elide causal order with temporal order, this is not correct, because the process matrix framework is specifically designed to study processes in the absence of a fixed background structure and so the notion of causal order that emerges out of that framework cannot simply be temporal order, although of course in real-life examples  we expect causal order and temporal order to coincide. Rather `causal order' here simply refers to the fact that in many processes, the operations occurring within that process can be arranged in a well-defined order which is a property of the process itself:  for example, if a process involves passing a single system between four laboratories, with an operation being performed on it in every laboratory, the order in which these operations take place is encoded in the process matrix and  can be inferred from the subsequent state of the system, since the net effect of performing the operations  $A , B , C , D$  (in that order) will in general be different from the net effect of  performing the operations  $A, C, B, D$ (in that order). Thus there is no need to appeal to the temporal ordering of the operations, and indeed it is at least logically possible that the order defined by the process matrix should fail to coincide with the temporal order (for example this could happen in a world where retrocausality is possible, or where there exist closed timelike curves). 

  The distinction  between causal order and ordinary cause-effect relations is less clear in the context of classical physics,   because classical operations  commute, and therefore in classical physics the order in which some set of operations are performed on a system $S$ makes a difference to the final state only if the choice of later operations depends on earlier operations in some way (for example, if we decide which operation to perform on $S$ conditional on the outcome of a measurement on $S$) - and in that case we can just say we have  cause-effect relations in the usual sense. So there don't seem to be many examples in classical physics where the order of operations is important in and of itself. On the other hand in quantum physics operations do not in general commute, and thus the order in which the operations are performed makes a difference to the final state, even if later operations do not causally depend on earlier operations. So in quantum physics it is particularly important to distinguish between causal order and cause-effect relations, because the former can be significant even when the latter are not present: we can have `pure' causal order instantiated simply by  a sequence of non-commuting operations without outcomes.

Thus we  propose a formal definition of causal order. First, we define the `signal requirement' as a condition on the strict partial order assigned to a set of laboratories which requires that for any possible choice of local operations by the agents in the laboratories, the outcomes obtained in each laboratory should depend only on  the settings in that laboratory or the settings for experiments which occur earlier in the order (note that this is the same requirement used in the definitions of causally ordered and causal processes, as described in section \ref{PM}). We recall that the word `depend' here refers to statistical dependency of the kind that could be used for signalling, hence the name `signal requirement.' Then we have the following definition:

\begin{definition} 
A subset $\{ X \}$  of the laboratories   $ \{ P \} $  belonging to a process $P$  exhibits causal order iff   there is some choice of strict partial order  $O$ for the remaining laboratories $\{ P \backslash X \} $ such that after conditioning on  the order $O$ for the laboratories in  $\{ P \backslash X \} $, a) there exists at least one choice of strict partial order for the laboratories in $X$ which is compatible with the signal requirement, and b) there exists at least choice of strict partial order for the laboratories in $X$ which is \emph{not} compatible with the  signal requirement. \end{definition} 

 The point of this definition is that for many  processes,   some  strict partial orders  are compatible with the signal requirement and some   are not, and thus  these processes can be regarded as having some inherent order, which is what we refer to as `causal order.' For example, a process composed simply of an error-free channel exhibits causal order because the only strict partial order compatible with the signal requirement is the one in which the laboratory containing the input to the channel is earlier than the laboratory containing the output to the channel. On the other hand, there are two obvious ways in which a process can fail to exhibit causal order (for simplicity, here we focus on processes composed of just two laboratories). First, a \emph{non-causal} process composed of two laboratories cannot exhibit causal order, because there  is \emph{no} choice of strict partial order for these laboratories which is compatible with the signal requirement. Second a causal process composed of two laboratories with no signalling from one lab to the other also cannot exhibit causal order, because \emph{any} choice of strict partial order  for these laboratories is compatible with the signal requirement.  More generally, a subset of laboratories in a given process can exhibit causal order only if the process allows signalling between at least two laboratories in that subset, since the signal requirement never places any constraints on the relative order of laboratories with no signalling between them. So a process can fail to exhibit causal order either because no possible order for it is compatible with the signal requirement, or because all possible orders for it are compatible with the signal requirement - either way, there is no `inherent order' which follows from the definition of the process together with the signal requirement.
 
One might worry that because this definition refers crucially to outcomes, it does not cover the motivating case of  pure causal order without outcomes. However, recall that according to the definition of a process, agents involved in the process should be free to perform any operations allowed by quantum mechanics, so we can always  consider an instance of the same process where the agents perform   operations which \emph{do} have outcomes, and then in general we will indeed be able to find some subset of the laboratories where the signal requirement places constraints on the strict partial order. Thus in a  sense, `causal order' is similar to `causal structure' in the context of Minkowski spacetime, since it expresses not standard cause-effect relations but some kind of higher order modal structure - it tells us where there \emph{could be} cause-effect relations if the operations required to obtain outcomes are carried out. We will have more to say about this connection later.

\subsection{Asymmetry \label{asym}} 

Is the observation that causal order plays an important role in quantum mechanics in   tension with the common  view that there is no causation in fundamental physics? To answer this question, we must examine more closely the reasons usually offered for this view. Russell has a number of arguments, but perhaps his most enduring point is that causation is usually supposed to be asymmetric  -  if $A$ causes $B$ then $B$ cannot also be the cause of $A$ - whereas the fundamental laws of physics  do not seem to be asymmetric or directed in any way: `\emph{The law makes no difference between past and future: the future``determines" the past in exactly the same sense in which the past ``determines" the future.}'\cite{10.2307/4543833} This point continues to appear in the work of current philosophers - for example, Frisch suggests that  the apparent time-symmetry of fundamental physical looks like a serious problem for causal fundamentalists, because  `\emph{the (fact that) the notion of causation is asymmetric ... may even be thought to constitute \textbf{the} core of our notion of cause,}'\cite{10.2307/25592014}\footnote{As we will see in section \ref{science}, Frisch subsequently offers a putative counterexample to the claim that fundamental physics is not in fact time-symmetric, and so argues that causal fundamentalism can be rescued.} and Field points out that this argument still succeeds even if the fundamental laws are not time-symmetric: all that is needed to make it is that the laws of physics `\emph{have essentially the same character in both (temporal) directions,}'\cite{Field2003-FIECIA} which does seem to be the case for most of our fundamental physical laws\footnote{Field does point out that certain interpretations of quantum mechanics seem to violate this criterion - most obviously collapse interpretations, since systems evolve differently after the collapse of the wavefunction. However we will not dwell on these cases here, as our purpose is to show that there can be a well-defined notion of causal order in fundamental physics even if it does not exhibit any directedness or asymmetries.}.  So  there is apparently nothing intrinsically directed  in fundamental physics which could play the role of  causation, and therefore it seems natural to conclude that there is no causation in fundamental physics - causation arises at a  macroscopic level, perhaps because of the thermodynamic gradient\cite{kutach2013causation} or due to the presence of agents who can play a role in  an interventionist\cite{pearl2009causality,woodward2005making} or perspectival\cite{Price2005-PRICP} analysis.

The presence of `causal order' in fundamental physics might seem like a counterexample to the claim that there is nothing intrinsically directed in fundamental physics, but in fact it  is not, because causal order does not require a notion of asymmetry or directedness. Here, a distinction must be made between   \emph{directedness} and \emph{order} - we can define an order on a set of objects which stand in relations to one another even if the relations are not intrinsically directed. Consider, for example, the rungs on a ladder which is perfectly symmetric, i.e. it looks the same if we flip the whole thing about an axis through the middle parallel to the rungs. Due to the symmetry there cannot be any `intrinsic direction' to the ladder, and so, given two rungs, there is no fact of the matter about which rung comes after the other; but nonetheless if we start at one end and go towards the other end without crossing any segment twice, we will end up with a well-defined total order for the rungs. This order is not the objectively correct order, and we could get another equivalent order by starting from the other end, but in either direction we get a well-defined total order.  In light of this distinction, in this article we will use the term  `bi-order' to refer to an order where the direction of the arrows does not matter: that is to say, a bi-order is the equivalence class consisting of an order and the corresponding order obtained by simply reversing the direction of all the relations making up the order. Thus, for example, a strict partial bi-order is a bi-order such that each of the two orders in the equivalence class is a strict partial order (note that reversing the direction of all of the relations in a strict partial order will always give rise to another strict partial order). 

We now observe that  `causal order' requires only a bi-order, not an ordinary order with an intrinsic direction. To see this, it will be helpful for us to use terminology due to Pienaar\cite{https://doi.org/10.48550/arxiv.1902.00129,Pienaar_2020} distinguishing between two ways of describing a process. First, the `observational scheme,' (OS) specifies the process by giving a joint probability distribution over the full set of variables, thus allowing us to  choose any set of variables and make predictions for the other variables conditional on the values of the variables in that set. Second,   the `interventionist scheme,' (IS) begins by choosing inputs and outputs and then specifies the process by giving probabilities for the outputs conditional on the inputs. The idea  is that the `observational scheme'  encodes only the fundamental physical relationships between variables that constitute the process, while the interventionist scheme consists of various possible descriptions of this process that come from embedding  it in  macroscopic scenarios where agents have control over certain variables and not others\footnote{The claim that the `observational scheme' is more fundamental than the `interventionist scheme' is tantamount to the idea that joint probability distributions over histories are more fundamental than conditional probabilities which predict one event in terms of previous ones; this idea has been discussed in another context by Wharton and Liu in ref \cite{https://doi.org/10.48550/arxiv.2206.02945}.}. So for example, if we have an error-free channel between variables $A$ and $B$, in the OS we would describe this process by the symmetric probability distribution $p(A = B) = 1$, whereas in the IS we could pick $A$ as the input and thus describe the process by the conditional distribution $\forall x \ p(B = x | A = x) = 1$, or we could pick  $B$ as the input and thus describe the process by the conditional distribution $\forall x \  p(A = x | B = x) = 1$ - notice that the IS descriptions look superficially asymmetric even though the underlying OS process postulates a perfectly symmetric relation between $A$ and $B$.

With this terminology in hand, let us now consider  a process in which agents in separate labs perform the non-commuting invertible quantum operations $A, B, C, D$  consecutively on a single quantum system. Nothing in the  fundamental physics of these operations singles out one particular direction: if we treat the input to $A$ as the initial state then we can write the resulting IS description in the form  $ABCD$, while if we treat the output of D as the initial state then we can write the resulting  IS description in the form $D^{-1} C^{-1} B^{-1} A^{-1}$, but if we throw away external information about time direction and the involvement of agents it would seem that there is no deep distinction between these two descriptions: that is to say, $ABCD$, and $D^{-1} C^{-1} B^{-1} A^{-1}$ can be regarded as two different IS  descriptions which correspond to the same OS joint probability distribution.  However, although the direction does not matter, evidently the effect of $ABCD$ is not the same as the effect of $ACBD$: we are free to write the operations in either direction in the IS, but since the operations don't commute,  whatever direction we choose is associated with a unique correct order, and if we change the order then the IS description will no longer correspond to the same OS process. In this sense this example exhibits causal order even though it is reversible, and thus  it  corresponds to a bi-order - in this case a total bi-order. 
 
What about the case where the agents perform operations that have nontrivial outcomes? Again, if the underlying physics is truly time-reversal invariant then in principle it would seem that given an IS description we should be   able to write a time-reversed IS description which will correspond to the same underlying OS process. This time-reversal will involve inverting  all the operations performed inside the individual laboratories, thus turning input variables into output variables and vice versa (see appendix \ref{app} for more details about time-reversal).   So suppose we have an IS description which exhibits causal order in the simplest sense, i.e. it has an output   $O_1$ which is correlated with an input $I_1$, so when we try to write down the `inherent order' of this process in accordance with the signal requirement, it must be the case that $I_1$ occurs before $O_1$.  Under time-reversal, $O_1$ becomes $I_2$ and $I_1$ becomes $O_2$,   and  if the reversal operation preserves the underlying OS joint probability distribution, then it must be the case that $O_2$ now depends on $I_2$. Moreover, if we simply flip the directions of all the arrows in the original strict partial order, this yields a new strict partial order according to which $I_2$ occurs earlier than $O_2$, so we obtain a new `inherent order' which remains compatible with the signal requirement. Thus  in at least some cases, reversing an IS description of a process which exhibits causal order will yield another process exhibiting causal order in the opposite direction, and this second process is  ex hypothesi an alternative description of the same underlying OS process.   This demonstrates that the definition of causal order does not require that there is any asymmetry built into the underlying physics:  to identify causal order we need only a strict partial bi-order rather than a strict partial order, since we can have processes compatible with two different IS descriptions for which the inherent orders associated with the signal requirement are oppositely directed.   And thus although the objections of Russell, Field and others provide   compelling reasons to think there are no ordinary relations of cause and effect in fundamental physics, they certainly do not rule out \emph{causal order} in the sense used in the process matrix framework, since that notion does not require any intrinsic asymmetry.

 But does this mean that the distinction between   processes which exhibit causal order and processes which do not is purely conventional, depending entirely on our choice to regard some variables as inputs and others as outputs? No, because the question at issue is not which ordering is correct, but   whether there is \emph{any} possible ordering. Thus we can   define causal order at the level of the OS description without specifying a choice of inputs and outputs at all - that is, we can say that an OS process exhibits causal order iff there exists at least one IS description of this process which exhibits causal order, whereas an OS process does not exhibit causal order if there is \emph{no} IS description of it which exhibits causal order. So there is a well-defined, non-conventional distinction between processes which exhibit causal order and processes which do not, even if all the processes we are dealing with   are symmetric and reversible.

\subsection{Causal order in physics \label{science}}

 We will henceforth refer to  cause-effect relations in the everyday sense as   `strong causation,' in contrast with the weaker notion of `causal order.'  We will not offer a precise definition of strong causation here; instead we suggest that readers should fill in the details according to their preferred analysis of causation - the idea is that strong causation captures whatever intuitions about causality one considers as essential to the concept, whereas causal order requires only to the presence of well-defined order and may not satisfy other intuitions about causality.   For example,  `strong causation' presumably requires  some form of asymmetry or intrinsic directedness,  and it has also been argued that causation should be deterministic\cite{Norton2003-NORCAF}, local\cite{Field2003-FIECIA,Norton2003-NORCAF}, or should involve only a small set of causally relevant factors\cite{Field2003-FIECIA}, so one might insist that strong causation should have these features. 
 
 This distinction now allows us to offer greater clarity on the meaning of the term `causal' in the terms  of \emph{causally ordered}, \emph{causally separable} and \emph{causal} in the process matrix formalism: since the signal requirement appears in the definitions for all of these terms, and meanwhile the definitions say nothing about asymmetry, determinism, locality etc, it is clear that these definitions should be understood as referring to  causal order rather than strong causation.  Moreover, having observed that causal order need not involve asymmetry, we can provide an answer to the problems for the process matrix formalism described in section \ref{probs}. For the concerns we raised there about the role of agents hold only if one is  determined to describe process matrices as encoding \emph{strong causation}; if instead we understand the object of study of the process matrix formalism to be \emph{causal order}, then we don't need to invoke interventionist or perspectival considerations to ensure that the causal structure has some asymmetry, and therefore we have no need to introduce any agents.

More generally, we suspect that equivocation between causal order and  strong causation is   the source of some of the ongoing disputes over the role of causality in fundamental physics, and thus distinguishing carefully between these concepts can resolve a number of outstanding questions.   For example,  Frisch\cite{doi:10.1093/bjps/axp029} argues that causal notions are intrinsic to dispersion relations, because the derivation of the dispersion relation relies on a condition that many physicists\cite{PhysRev.104.1760, 1972causality} describe as `causality,' stipulating that `\emph{no output can occur before the input.}'  As Frisch notes, there is `\emph{no purely formal or mathematical reason}' for this requirement but it is nonetheless accepted as physically well-founded and indeed treated as `\emph{a general constraint on all physically plausible models}.' The causal ordering account explains why, for this `causality' condition is essentially a version of what we have called   `the signal requirement' and thus there is a very good physical reason why it should be obeyed: because to the best of our current knowledge, all physical processes are causal (and arguably they are required to be so in order to avoid logical contradictions - see section \ref{why}). Thus although  the causality condition used here is superficially time-asymmetric, as Frisch emphasizes, nonetheless it does not indicate the presence of any intrinsic asymmetry, as we have seen that well-defined causal order does not depend on asymmetry. So Frisch's observation that causal notions play an important role in the dispersion relations is not in tension with the fact that the physics involved in these relations seems to be time-symmetric, since it is causal order and not strong causation that is relevant here.

Similarly, although physicists very frequently refer to `causal structure' in the context of special relativity, there has been controversy over whether special relativity really has anything to do with causation: for example,  Nerlich\cite{10.2307/687169} argued  that causality plays no fundamental role in the theory, for after all the lightcone structure can be identified purely in terms of the behaviour of light, without any reference to `causation.' Nerlich also demonstrated that attempts by Zeeman\cite{doi:10.1063/1.1704140}, Malament\cite{malament1977causal} and Winnie\cite{Winnie1977-WINTCT}  to construct special relativity on the basis of a causal relation fail to show convincingly that the relation in question is really `causal' in any strong sense. But this controversy can be resolved by appeal to the distinction  we have just made: lightcone structure in special relativity defines \emph{causal order}, because it ensures that spacetime points are  arranged in a strict partial order or bi-order which defines the possibilities for signalling between them,  but it does not necessarily say anything about \emph{strong causation}. To have strong causation   we need some kind of asymmetry, and we will not get this from special relativity alone, since the time-reverse of a special-relativistic spacetime is another valid special-relativistic spacetime. So spacetime structure is genuinely causal in the sense of `causal order,' but it does not instantiate \emph{strong} causation and in this sense Nerlich's criticisms are well grounded. 

Furthermore, notice that spacetime structure and causal order as we have defined it are  structurally similar:   neither one really depends on asymmetry or directedness, and both ultimately instantiate a strict partial order (which is really a bi-order). And indeed, the connection between them has often been  recognised.   One possible approach to that connection is explored in ref \cite{renato}, which  sets up a framework to study the relationship between `causality' in the spacetime structure sense and `causal order' in the process matrix sense. This approach involves embedding a causal structure defined using the process matrix formalism (or another similar information-theoretic formalism) in a spacetime, and then imposing compatibility constraints - for example, a system should only be able to signal to other systems in its future lightcone - so that ordering requirements for processes  can be understood as consequences of relativistic causality. An alternative approach  is taken by the causal set programme\cite{Sorkingeometry}, which  builds spacetime up out of a set of causally related events which are required to stand in a strict partial order.   From this point of view, rather than trying to embed a causally ordered structure in a spacetime, we can  simply allow the spacetime to emerge from the  causally ordered structure. Moreover, the methodology of the process matrix formalism adds a new nuance to this construction,   because it allows us to see that the `strict partial order' requirement need not be imposed by fiat but instead could be derived from  the need to avoid contradictions (see section \ref{why}), and  thus from this point of view, spacetime structure  could potentially be regarded as a consequence of basic consistency requirements for valid processes.

 \section{The Metaphysical Project: Strong causation versus causal order } 
 
 We now move to the metaphysical project, where we will argue that the notion of causal order identified in the previous section offers new insight into the nature of causation.  For although causal order does not fully satisfy our everyday intuitions around causality, nonetheless   it clearly bears some relation to strong causation.  Indeed,   once we select a preferred direction along the strict partial bi-order associated with  a process exhibiting causal order, it then becomes an ordinary strict partial order, which can be interpreted as encoding strong causation as in the causal modelling framework. Moreover, in many cases there is indeed a natural choice of direction along the bi-order arising from the particular perspective that we have on it - for example, usually we experience ourselves as having control over certain variables and not others, so it is natural to regard the controllable variables as `inputs,' and as we have seen, specifying a choice of inputs for a causal process singles out a choice of direction for the associated bi-order. 

In this connection, it is important to emphasize that choosing a direction for a bi-ordered graph is not the same as adding a direction to an undirected graph. If you simply take an undirected graph and assign a direction to one of those edges, that choice does not  define a direction for all the other edges; you are free to assign directions to the other edges in many different ways, and in a  sufficiently complex graph it will usually be possible to do this in such a way that the resulting directed graph fails to be acyclic. Whereas in a bi-ordered graph there is a well-defined `flow' through the graph, although the direction of the flow is immaterial; so there are exactly two valid ways of assigning directions to the edges, and one is the reverse of the other, such that choosing a direction for any one edge is equivalent to selecting one out of the two valid  assignations of directions. Therefore  in this case we are not free to simply create loops: if the bi-order is acyclic, then any choice of direction will lead to an acyclic graph. So it is specifically the fact that the underlying process exhibits causal order which ensures that invoking   interventionist or perspectival considerations  to select a direction leads us naturally to a structure which can be represented by an acyclic causal model and which can therefore be regarded as encoding strong causation. This idea is explored in more detail in ref \cite{pittphilsci20539}.

Indeed, causal order is clearly crucial to any account of strong causation, although this point is not often articulated in accounts of the emergence of macroscopic causality.  Invoking a perspective or an intervening agent provides a local direction of causation, but this will not yield   well-defined macroscopic causal structures more globally unless there is some underlying causal order that ensures consistency of causal direction throughout the network of related events. Often the importance of causal order goes unmentioned because there is an implicit assumption that we are working on a fixed spacetime background, so temporal order can be substituted for causal order  - but in fact, this is not enough to guarantee a well-defined causal structure unless retrocausality is ruled out by fiat, and if the aim is to account for the emergence of causality it is hard to achieve this in a non-question-begging way. So to arrive at a non-circular account of strong causation we must start from a process which exhibits some form of causal order, and thus when we add perspectival or interventionist considerations we will arrive at a  well-defined global causal structure without loops or other pathologies.\footnote{  Of course, adding a direction to the causal order associated with quantum mechanical systems will only yield strong causation in the context of the description of  certain sorts of highly specific quantum experiments; to move from fundamental physics to the kinds of causal laws that feature in our everyday life, like `Exercise prevents heart disease,' would be an extremely complex problem requiring us to deal with various confounding factors and to settle the question of how to reduce the classical to the quantum. So for most instances of strong causation in the macroscopic world there is no straightforward way to explicitly derive the causal laws from quantum causal order;  but nonetheless it is clear that we would not be able to arrive at  macroscopic causal laws if it were not the case that macroscopic events exhibit some kind of causal order, and it seems reasonable to think that the causal order of the macroscopic world has its origin in the fact that  certain sorts of processes in the underlying microscopic physics also exhibit causal order, even if we can't at present perform that derivation explicitly.}

 \subsection{Causal Modelling \label{cm}}
 
This claim  about the origins of causality has consequences for the causal modelling programme in quantum foundations.  For we have emphasized that no-signalling processes cannot exhibit causal order\footnote{It should be emphasized that this claim about no-signalling processes applies only to \emph{genuinely} no-signalling processes, like the Bell correlations: there is nothing anyone can do to make the Bell correlations transmit a signal, so they are simply not subject to ordering requirements. On the other hand, in cases where we have `no-signalling' only because two classical causal influences  cancel out, a small change in the initial conditions will turn the no-signalling process into a signalling one, and thus we would expect that these sorts of influences will still be required to stand in a well-defined order, which in turn justifies the original instinct to refer to them as `causal.'}, so if  it is accepted that causal descriptions are typically grounded by underlying causal order, it seems that   one should \emph{not} expect that causal notions will be a suitable way of analysing no-signalling processes. And in particular, since measurements on entangled quantum particles are no-signalling, we shouldn't expect to be able to find a natural causal account of  phenomena involving measurements on entangled particles, such as the Bell correlations. Of course there is nothing to stop us from writing down causal models for such processes, but it's unclear that such models  can be  regarded as describing any real feature of the processes, since ex hypothesi causal phenomena arise from  causal order and there is no causal order involved here.

Moreover, this conclusion exactly matches what we observe in reality: no-signalling correlations from quantum entanglement can be exhibited between measurements at spacelike separation, in which case there is no natural temporal order to assign to them, whereas signalling processes can never be performed at spacelike separations, and therefore they are always associated with a temporal order. That is, in quantum mechanics we find that signalling processes are always ordered (temporally) whereas no-signalling processes don't usually seem to  have any inherent order, which is what we would expect based on our argument that no-signalling processes do not exhibit causal order. 

  This observation raises questions about certain conclusions that have been drawn using causal models in quantum foundations. For example,  the fact that any causal model which reproduces the Bell correlations must be fine-tuned is sometimes taken to mean that there is some mystery about the Bell correlations which urgently requires explanation\cite{SpekkensWood}, but the line of reasoning we have pursued here tells us that it's not surprising that we are unable to find a satisfying causal model for the Bell correlations, because the Bell correlations are simply not the right kind of thing to analyse with a causal model.  Note that this is true even though the Bell correlations can  be described in interventionist terms, with the preparation and measurements playing the role of the `loci of possible interventions' - an interventionist analysis can succeed in transforming a bi-order into a directed acyclic graph only if there is some well-defined bi-order present, so in the case of the Bell correlations, since  these processes need not instantiate causal order, there is no reason to expect that an interventionist analysis will yield a sensible causal model. Of course, results like  that of Wood and Spekkens are still very important to help us understand the specific ways in which standard causal models fail in the context of quantum mechanics, but the point is that we should not be too distressed to find that no non-fine-tuned causal model is possible for these phenomena.

  In response to this argument, proponents of the causal modelling approach could of course reject the account we have suggested here of the origins of causal phenomena, and maintain that  there  exists a set of  ontic variables underlying quantum mechanics which stand in genuinely asymmetric causal relations, so strong causation is in fact a feature of reality at the most fundamental level. However, we would then be required to accept the following story: 
  
  \begin{quote} \emph{Nature contains asymmetric causal relations at the most fundamental level. This  fundamental causal structure gives rise to quantum mechanics, which is time-symmetric\footnote{As shown recently in ref \cite{dibiagio2020quantum} quantum mechanics is time-symmetric even if we include the Born rule.}, so the underlying asymmetries have  become invisible. Then from this temporally symmetric physics there   arises a higher-level macroscopic description which contains asymmetric causal relations, but these macroscopic asymmetries have nothing to do with the fundamental asymmetries - they may perhaps be explained as a consequence of the second law of thermodynamics, or simply given a perspectival account based on the fact that agents perceive themselves as having some temporal orientation, but crucially, at no point do  fundamental asymmetric causal relations play any part in this story.  So the causal features of the macroscopic world are not connected   in any way to the causal structure that underlies quantum mechanics.} \end{quote} 
  
  The  problem with this story is that if macroscopic causal features have nothing to do with the putative underlying causal structure, then macroscopic causal features cannot count as evidence for the existence of that underlying causal structure, so it's  unclear why we should believe in  this putative underlying causal structure. After all,  ex hypothesi these fundamental asymmetric causal relations impact macroscopic reality only via the role they play in giving rise to  time-symmetric quantum mechanics, so macroscopic reality would exhibit exactly the same causal features even if there were no underlying asymmetries. Indeed, it would surely be quite a strange coincidence were the causal asymmetries of the macroscopic world to be reflected so precisely in an underlying fundamental asymmetry, even though this fundamental asymmetry apparently has nothing whatsoever to do with the asymmetry of the macroscopic world - is it not more likely that this hypothesis is simply the result of erroneously extrapolating the causal structure of our macroscopic experience?

To push back against this criticism, the proponents of the causal modelling framework might observe that  at least for the moment it seems that  not all possible process matrices actually occur in reality - at present we only know how to implement the causal ones. Therefore one might be tempted to argue that the reason we see only causal process matrices is that reality is fundamentally `causal' in the sense of strong causation, and thus the only processes which can occur are the ones which have an underlying causal structure. This would seem to justify the methodology of seeking to come up with an underlying causal structure for the quantum mechanics, including the Bell correlations. However, it remains possible that the non-causal processes are in fact physically possible, and simply occur only in special regimes,  such as in quantum gravity\cite{Hardy_2009} or on closed timelike curves\cite{Oreshkov2} (after all,  CTCs do in fact occur in some solutions to the Einstein Field Equations\cite{universe7010012}). Moreover, even if it is true   that only causal processes occur in nature, it remains the case that there doesn't  appear to be  intrinsic asymmetry in these processes, so there is no reason to invoke specifically \emph{strong} causation to explain the absence of non-causal processes. At best we have reason to hypothesize that nature is fundamentally causal in the sense that all physically possible processes can be assigned a strict partial order which obeys the signal requirement, and since the signal requirement places no restrictions on no-signalling processes, \emph{this} hypothesis does not justify us in seeking to come up with an underlying causal structure for the Bell correlations.

\subsection{Why causal order? \label{why}}

In addition to offering a novel account of the origins of causal phenomena, the process matrix framework takes a further intriguing step: it offers a possible explanation for the \emph{necessity} of causal phenomena. For the whole point of the process matrix framework   is that it is defined in such a way as to avoid contradictions, and therefore it directly incorporates a set of consistency conditions on dynamics which must be obeyed if agents locally have the freedom to perform operations and compose processes in any way they like. And as we have just noted, this places strong limits on processes which are  signalling. In particular,  a large class of process matrices (indeed, all those process matrices known to be physically possible) avoid contradictions specifically because they are causal, meaning that there is at least one choice of strict partial order for the laboratories involved which obeys the signal requirement: evidently this rules out scenarios like the one depicted in figure \ref{signalling}, where signalling processes are used to give rise to a contradiction.  Moreover, any causal processes which allow signalling between different laboratories  will necessarily instantiate causal order, and thus the consistency conditions encoded in the process matrix formalism  help to explain the phenomenon of causal order.

 On the other hand, we have emphasized that causal order cannot be instantiated by laboratories  which  can't signal to one another. This makes sense if we think of the signal requirement as a kind of consistency condition, because no-signalling processes cannot give rise to contradictions in the style of figure \ref{signalling}, and therefore there is  no need for no-signalling processes to  have any inherent order.   This underlines our previous point about the inappropriateness of applying causal modelling to no-signalling correlations: not only is it the case that no-signalling correlations can't exhibit causal order, they are not even subject to the kinds of consistency requirements which one might regard as the underlying reason for causal order.

Thus the consistency conditions encoded in the process matrix formalism offer a partial explanation for the prominence of causal order in the physical world, as well as the apparent absence of causal order for processes like measurements on entangled particles. However it is only a partial explanation, because we can also use the process matrix formalism to write down  \emph{non-causal} processes which obey the consistency conditions,  so something more than just consistency conditions is needed to explain why it seems that all physical processes are causal, or at least why  the regimes we have so far probed contain only causal processes. For example, ref \cite{Ara_jo_2017} proposes that all physical processes should be `purifiable,' i.e. they should be expressible as part of a pure process in a larger space, where a pure process is one that  induces a unitary transformation from the past to the future whenever unitary transformations are also applied in the local laboratories. It is shown that this rules out a large class of non-causal processes, although not all of them. Several other arguments are presented in ref \cite{pittphilsci20539}. First, if we allow non-causal processes, then the dynamics of the allowed processes cannot be defined in a local way, since certain sorts of dynamics  are possible outside closed timelike curves and exotic causal structures but are not always possible inside such structures - e.g. we saw in section \ref{PM} that in the classical process function approach the identity operation around a closed timelike curve is not possible, but evidently the identity operation is possible outside closed curves.    One might argue that this novel kind of nonlocal dependence on the global structure of spacetime is not permitted by the laws of our actual world, and thus non-causal processes are not physically possible. Second,  the notion that non-causal processes are possible relies crucially on the idea that there is a fundamental distinction between local operations in laboratories, which can be freely chosen, and dynamics between the laboratories, which are fundamentally uncontrollable - for example, the framework must have the consequence that if we try to implement the identity operation around a closed curve we will always fail.   And yet in a sense one can think of dynamics as being just a sequence of chained local operations, so one might feel that the distinction being made here is too strong: if we have free choice over local operations, we should also be able to freely construct dynamics on a CTC by means of chains of local operations. This line of reasoning might be taken to suggest that  these sorts of non-causal processes are not in fact physically possible - or at least they should not be possible in the kinds of regimes where we can in principle have very good control over the dynamics we implement, which would explain the fact that we have not observed any of them in the regimes we have thus far probed. In any case, if some such argument is accepted, we can then argue that the necessity for all signalling processes to exhibit causal order follows directly from the need to avoid contradictions, and thus the causally ordered nature of the physical world can be regarded as a simple consequence of basic consistency conditions.

This way of thinking allows us to bring some additional nuance  to the approach to causality adopted in the process matrix framework. As we have noted, this approach largely identifies causal influences with signalling, or at least uses signalling as a proxy for causation, whereas  most people consider that signalling is at best a sufficient condition for the presence of a causal relation\cite{renato}.  However, the understanding that causal ordering is related to consistency requirements allows us to avoid presupposing the association of causation with signalling, and rather to derive it from more fundamental considerations. That is, suppose we accept that ordinary causal phenomena arise from an underlying causal order together with a choice of direction. Then recall that consistency requirements impose conditions only on signalling processes, not on no-signalling ones. It follows that causal order, and hence causal phenomena,  will typically be associated with signalling processes and not with no-signalling ones, so the association of `causation' with `signalling' in this framework can thus be retrospectively justified.

\subsection{Modal Structure \label{ms}}

One might use the considerations we have presented here to argue for a kind of causal fundamentalism, where one would argue that the phenomenon of causal order demonstrates that there  \emph{is} in fact  causation in fundamental physics, even if it does not involve any temporal asymmetries. Ney advocates a position of this kind in ref \cite{10.2307/25592033}, suggesting that at the fundamental level, `\emph{There is still causation, because there is still physical determination. But the distinction between what is the cause and what is the effect may not be fundamental.}' That is to say, she holds that we can identify causation with fundamental symmetric relationships  and allow that the asymmetry of causation only  emerges at a higher level. Evidently Ney's take on causal fundamentalism  fits nicely with the picture we have advocated here, where causal order exists at the fundamental level but causal asymmetry appears only when we use higher-level considerations to select a direction through the bi-order. 

Alternatively, one could parse our approach as a  causal non-fundamentalist viewpoint, where the causally ordered nature of underlying fundamental processes is not itself regarded as a form of causation but instead is taken to be a partial truth-maker or grounds for higher-level causal claims. In this sense, our account could be regarded as one possible way of realising the vision set out by Norton, who argued that  `\emph{a causal character can be recovered from the science as looser, folk sciences that obtain in restricted domains}' - one could interpret this as a description of the way  in which strong causation emerges from causal order in special domains where there is a natural way of choosing a direction for the bi-order. However, we emphasize that since causal order is essentially a kind of modal constraint (determining which processes can and cannot exist, and how they can and cannot be arranged in spacetime) this position still requires a kind of \emph{modal} fundamentalism - that is to say, it does not   eliminate objective modality altogether, but rather reduces causation to a deeper modal structure which is more consistent with  known physical law.

The idea that fundamental physics involves causal order but not strong causation may raise some concerns amongst the subset of scientific realists who argue that realism is necessarily requires causal explanation\cite{schmid2020unscrambling,SpekkensWood}. This kind of position seems to be related to  the common idea that there is a contrast between (bare) correlation and causation, as epitomized by the slogan `correlation is not causation,' which comes with the implication that `causation' expresses something stronger than bare correlation. And indeed, if `(strong) causation' and `bare correlation' were our only options for describing the character of fundamental physics, it would seem reasonable to say that the realist must be committed to a causal description,  since committing only to `bare correlations' would lead to a collapse into constructive empiricism\cite{van1980scientific}.  However,  (strong) causation and bare correlations are not in fact the only possible options. In particular, we would   argue that   causation is stronger than correlation not because it is \emph{asymmetric} (after all, bare correlations themselves can also be asymmetric) but  because it is \emph{modal}. It is precisely this modal force which is captured in the interventionist account of the difference between causation and correlation -  `\emph{only some and not all correlational relationships are potentially exploitable for purposes of manipulation and control; we regard those relationships that are so exploitable as causal,'}\cite{Woodward2014} - since modal relations are required  to ground counterfactuals describing what would be the outcome of various kinds of manipulation and control.  So we can still give a realist account of fundamental physics without invoking strong causation, provided that we are willing to make some other kind of modal claim, for example by postulating \emph{symmetric} modal relations.

What exactly does a `symmetric modal relation' look like? Ney bases her account on a symmetric modal relation that she refers to as \emph{determination}, defined as `law-governed co-instantiation,' which could be expressed quantitatively in the form $p(A = B) = 1$. More generally, we could also imagine other  `law-governed' modal relations expressed in  the form $p(A = B) = x$ for any $x \in (0, 1)$. These sorts of modal relations are  less familiar to us than cause-effect relations,  as evidenced by the fact that there does not seem to be any word for them in ordinary non-technical language, but they are in fact ubiquitous in science. For in Cartwright's terminology, many laws that appear in science are not causal laws but rather `\emph{laws of association  ... (which) tell how often two qualities or quantities are co-associated}'\cite{10.2307/2215337} - and indeed this observation was indeed another of Russell's arguments for his claim that causation does not appear in fundamental science\cite{10.2307/4543833}. For example, consider a relation like the Maxwell-Faraday equation, $ \nabla \times E = -\frac{ \partial B}{\partial t}$. This equation is not directed: it is not the case that the state of the electric field \emph{causes} the magnetic field to have a certain rate of change or vice versa, since there is no fact of the matter about which `comes first.' But nonetheless the relation between $E$ and $\frac{ \partial B}{\partial t}$ is not just `bare correlation' - it is an instance of Ney's `law-governed co-instantiation,' which is to say it is nomically necessary that the two must stand in this symmetric relationship. Moreover, although we tend to think of dynamical laws as being intrinsically directed, Ismael emphasizes that dynamical equations are not really different in form from these co-instantiation laws: `\emph{Dynamical laws ... are constraints on the relationships between states at different times, but ... there is nothing in the law itself to say either determines the other.}'\cite{ismael2016physics} So in fact, there are good reasons to suppose that we should think of dynamical equations in the same way as equations like the Maxwell-Faraday equation - they simply express a symmetric, undirected modal relationship  between states at different times.  Thus realism need not be tied to causal explanation, as we can offer   a legitimately realist account of fundamental physics which replaces causation with symmetric modal relations linking events at different times.

These observations  allow us to bring additional nuance to our argument that the Bell correlations should not be given a causal explanation, for we can now emphasize that this does not mean they must be regarded as simply   `bare correlations.'  Instead we can say that the Bell correlations are ultimately to be explained by symmetric modal relations, just like the relations involved in ordinary laws of association and arguably even dynamical laws. The only difference between the Bell correlations and more ordinary cases of signalling correlations is that the signalling correlations  are subject to consistency conditions which require that they must have some inherent order (or rather, bi-order), but the bi-order requirement is not the explanation of the signalling correlations -  it only explains  the way in which the events that instantiate them are arranged. In both the signalling and the no-signalling case the correlations are ultimately explained by appeal to underlying symmetric modal relations, so the Bell correlations are just as explicable as signalling correlations.

 \section{The Functional Project: Non-Classical Causal Models\label{xx}}
 
 We have reached a broadly negative verdict on the applicability of classical causal modelling to quantum mechanics based on the observation that  the usual grounds for a causal description is not present in all quantum processes.  But one might still hope to offer a justification for causal modelling in quantum mechanics in the context of the functional project, which, in Woodward's words, `\emph{takes as its point of departure the idea that causal information and reasoning are sometimes useful or functional in the sense of serving various goals and purposes that we have.}' With this motivation in mind, one might argue that  causal modelling in quantum makes sense because it can serve an important functional role in the description of quantum phenomena, even if the  usual physical underpinnings of causation are absent. 

In this context, we note Woodward's warning that  `\emph{“cause” might not turn out to be like “entropy” in the sense that it is a notion that it is usefully applicable to physical situations with a certain structure when analyzed at a certain level of description, and not usefully applicable elsewhere,}'\cite{10.1086/678313}. This suggests a possible objection to the functional programme - one might worry that this `certain structure' which makes causal notions `usefully applicable' is simply \emph{causal} structure, in which case the functional analysis of causation apparently cannot be given in a non-circular way. But the approach we have taken here suggests a natural answer to this objection: the relevant kind of structure is  in fact \emph{causal order}, since the presence of causal order entails that there is a uniquely correct way to understand the situation in causal terms once we  select a direction for the underlying bi-order. Thus the functional analysis can be pursued without risk of circularity if we regard it as an  analysis of \emph{strong causation} and we understand the requisite underlying structure in terms of the independent concept of causal order.  And if this is accepted, it seems   that we should not expect causal models in the traditional sense to be `usefully applicable' in the description of no-signalling quantum processes, since these phenomena do not have the right kind of structure. 

However, this argument does not rule out the possibility that there could be some kind of generalized `quasi-causal' description which might play a similar kind of functional role to a traditional causal model in the context of phenomena which fail to exhibit causal order. Indeed, insofar as these phenomena are to some degree under the control of macroscopic agents,  it must surely be possible to arrive at a  description  of them which characterises the way in which they can be used to achieve practical goals and purposes, just as a traditional causal description would do. And this is exactly what is achieved by the formalism of quantum causal models, as presented in section \ref{causalmodel}. For quantum causal models do not appear to be attempting to describe an `underlying reality'  for quantum mechanics, but rather the reverse - they provide a `top-down' description  of the laboratory procedures that are implemented on the relevant systems, so for example ref \cite{schmid2020unscrambling} suggests that quantum causal models are suitable for the case where `\emph{one typically does not have a direct description of the causal mechanisms, but rather only a very coarse-grained description of them in terms of laboratory procedures that are implemented on the relevant systems}' and ref \cite{https://doi.org/10.48550/arxiv.1906.10726} notes that if this methodology is applied to characterise a quantum circuit, it  will often simply yield the connectivity graph representing the way the gates are literally wired together in the laboratory\footnote{Or it may yield a subgraph of this connectivity graph, because there may turn out to be nodes which do not influence one another even though they are physically connected.}. Essentially, the quantum causal structure provides a set of instructions for  the minimum possible laboratory operations which would be needed to produce the desired effects. Thus these models serve a similar function to causal descriptions in that they show us  how to construct quantum circuits which are suitable to achieve various goals,  so from the   functional point of view  the idea that these models have a claim to be thought of as   quasi-causal seems very reasonable.

 However, the literature also  contains some suggestions about the foundational significance of these models which might be interpreted as going beyond these operational claims. For example, ref   \cite{https://doi.org/10.48550/arxiv.1906.10726} suggests these models can be used `\emph{to establish an account of causality in quantum theory’s own terms, without assuming a separate realm of classical systems or measurement outcomes. The account provides its own answer to many of the basic questions concerning causality in a quantum universe.}' The idea is that if we    redefine `cause' in the quantum regime to refer to quantum causal models rather than classical ones, then we can after all have `\emph{causal explanations of quantum processes}'\cite{Barrett_2021}. But some caution is required here -  the fact that `quantum causal structure' offers a practically useful way of describing certain laboratory phenomena does not necessarily entail that it is providing a meaningful   \emph{explanation} of those phenomena. After all, `quantum causal structure' is not causation in the traditional sense, so even if one takes it that a traditional causal account is automatically explanatory, something more must be done before we can accept that this kind of structure is similarly explanatory. And  since it is a truism that it is possible to give an operational description of laboratory procedures,  the existence of such a description for some phenomenon should not be taken as revealing any deep fact about the phenomenon beyond the fact that it can be implemented in a laboratory.  In particular, we  caution that the use of quasi-causal concepts in operational descriptions of phenomena like the Bell correlations should not be allowed to obscure the very important point that the Bell correlations and other nonlocal phenomena do not exhibit causal order,  which as we have seen in this article can be regarded as an important clue about the right way to understand the Bell correlations and how they relate to spacetime causal structure. This observation is entirely consistent with the claim that the Bell correlations can be embedded in a top-down `quantum causal model' describing the operational procedures needed to produce them:  the point is simply that this kind of operational description does not necessarily tell us anything about the nature of the Bell correlations. 

 Of course, proponents of quantum causal models might be inclined to insist on functional grounds that there is actually no distinction  between a `causal explanation' and the kind of operational account offered by quantum causal models. We have no particular objection to this position, but we emphasize that taking such a stance would have the consequence that   a `causal explanation' for quantum phenomena would no longer confer any particularly novel  insight into those phenomena. For example, what the quantum causal modelling framework tells us about the Bell correlations is simply  what we already knew, i.e. that these correlations can be produced in the lab by an agent who prepares a joint entangled state in the causal past of both measurements. There may be good functional reasons to refer to this description as a `(quantum) causal model' but the name alone does not confer on the description any additional explanatory power - for example, those who previously believed that the Bell correlations must be given a local account by means of retrocausality or superdeterminism will surely not be swayed from their view by  linguistic or metaphysical arguments concerning what counts as a `cause.' \footnote{On a different note, in light of the preceding discussion it is interesting to consider the symmetry properties of quantum causal structures.  Quantum causal models replace causal relations with unitary transformations, and as noted in ref  \cite{https://doi.org/10.48550/arxiv.1906.10726} `\emph{the causal structure of a unitary transformation is reversible in a strong sense, which goes beyond the obvious fact that unitaries are reversible transformations: if the causal structure of a unitary transformation U is represented by drawing arrows from inputs to outputs ... then the causal structure of $U^{-1}$ is given by inverting the arrows. In case this seems obvious, note that a similar thing is not true for classical reversible functions.}' So `\emph{any temporal asymmetry ... in the framework of quantum causal models does not arise in the specification of the causal relations themselves}'\cite{https://doi.org/10.48550/arxiv.1906.10726}.  Hence the quantum causal modelling framework in fact contains no intrinsically asymmetric `cause-effect relations,' but instead instantiates   exactly the notion of causal order that we have set out in this article, which may be regarded as further evidence that the physics of quantum mechanics instantiates causal order but not strong causation.}

\subsection{Other Nonclassical Causal Theories} 

In light of the operational character of existing quantum causal models, there have been proposals that we should seek more radical versions of nonclassical causal theories to characterise the reality underlying  quantum mechanics. For example, ref   \cite{schmid2020unscrambling}  sets out to find a way of unscrambling the purely inferential features of quantum mechanics and the features which `\emph{describe realities of nature,}' where the latter are assumed to be described by a `realist causal theory.' Here it is argued that classical causal models aren't viable because of fine-tuning issues and no-go theorems, and that quantum causal models    are not sufficiently realist, so we need to find some other kind of nonclassical causal theory to represent the realities of nature. Moreover, ref \cite{schmid2020unscrambling} places great emphasis on the idea that this representation must still be sufficiently causal, noting that `\emph{Any attempt to provide a nonclassical generalization of the notions of causation and inference, however, is highly constrained insofar as it will need to preserve those features of these notions which one judges to be essential}' and arguing that the nonclassical theory should contain `\emph{analogues of most, if not all, of the standard notions that arise in the framework of classical causal models: common causes, causal mediaries, d-separation, evaluation of counterfactuals.}'

However, given the position we have set out here, the justification for seeking a specifically \emph{causal} model to represent the realities underlying quantum mechanics seems somewhat unclear.  After all,  if one agrees that `causality' in the usual sense arises from causal order plus a choice of direction, then not only is there no reason to expect a classical causal model to work in the context of the Bell correlations and other nonlocal quantum phenomena, there is no obvious reason to expect any other kind of causal description to work either - the conditions which typically give rise to the appearance of causal phenomena are  absent, and in their absence why should we expect that the underlying reality should have causal features at all?    Thus this emphasis on explaining quantum mechanics in terms of a `nonclassical theory of causation' seems too limiting: for example,  both classical and quantum causal models are required to obey a Markov condition, so presumably any further `nonclassical theory of causation' would also obey such a condition, but  if we are really seeking to understand the reality underlying quantum mechanics it's not very clear why we should restrict our attention to models with this feature, and indeed several authors have argued that realistic models for quantum mechanics should \emph{not} obey a Markov condition\cite{2008PhRvA..77b2104M,Adlamspooky}.  Similar points can be made for other features of causation which ref  \cite{schmid2020unscrambling} would wish to preserve in a nonclassical causal theory; while it could certainly turn out to be the case that the reality underlying quantum mechanics does in fact have some of these   features, and one might reasonably decide pragmatically to begin the search for possible models by looking at models with these features, nonetheless we should not rule out other possibilities just because they don't look sufficiently `causal' to our classical intuition.

Similarly, ref  \cite{renato} argues that `\emph{in light of Bell experiments, we are forced to update our prior understanding of the interface between information-theoretic and spacetime notions of causality and events,}' and thus  sets out  to present a framework which allows us to disentangle the information-theoretic and spacetime notions and reconnect them more carefully. However, the line of argument pursued in this article suggests that  some quantum phenomena should not be associated with an information-theoretic notion of causality at all - that is simply the wrong way to describe them.

Another interesting line of research involves extending quantum causal models to allow for cyclic causal structures\cite{Barrett_2021,https://doi.org/10.48550/arxiv.2109.12128}. In light of the arguments we have made here, this seems like a promising approach, since one way to model symmetric relationships which can't be decomposed into a strict partial order would  be to write them as cyclic structures in the context of a broader causal description. However, we would still caution that the expectation that the underlying structure should be \emph{causal} (albeit not acyclic) should be treated with care, since these sorts of phenomena do not   involve causal order and therefore there is no obvious reason to expect them to be best explicable in the context of a causal   formalism, even one that has been generalized in certain ways. For example, ref \cite{https://doi.org/10.48550/arxiv.2109.12128} continues to maintain that  the Bell correlations should be understood in terms of `\emph{causation without signalling (i.e., fine-tuned causal influences)}' whereas we have argued here that this is not the right way to think of the Bell correlations. This does not mean that research on cyclic causal structures  should not be pursued, but it does mean that we should be prepared to be quite flexible about these kinds of models: assumptions that they must have certain prototypically causal features should be carefully examined, and discarded if they are not working well.

 \subsection{Possible defences of the causal approach \label{defence}}

In support of the  idea that the `realities of nature' should be described by a (suitably generalised) causal theory, it has been argued  that the purpose of quantum foundations is  to provide an interpretation for quantum physics which retains as many concepts from classical physics as possible, so we should aim to retain the common-sense notion of causation. For example,  ref \cite{https://doi.org/10.48550/arxiv.1906.10726} motivates its use of causal modelling for quantum mechanics with the observation, `\emph{Reasoning in causal terms is omnipresent, from fundamental physics to medicine, social sciences and economics, and in everyday life.}' However, before accepting this argument we should first try to understand what reasons we might have for aiming to  retain concepts from classical physics in our interpretation of quantum mechanics. One possible answer is that we do so purely on pragmatic grounds, because we are unable to understand interpretations which don't retain such concepts. And indeed, it may well be the case that we are cognitively incapable of grasping concepts which are too distant from the classical reality to which our brains are adapted, but this doesn't really seem like a danger in this instance: the notions  of symmetrical modal relations and causal order may be unfamiliar but they are surely not incomprehensible. Another possible answer is that we should try to retain concepts from classical physics in our quantum interpretations in order to help explain why these concepts appear in the classical limit. But  this kind of argument is  valid only if the presence of the relevant concept in the   interpretation of quantum mechanics actually plays some role in explaining the appearance of the corresponding concept in the classical limit, and that does not  seem to be the case here: as we observed in section \ref{cm}, the existence of asymmetric relations underlying quantum mechanics would apparently be completely unrelated to the emergence of macroscopic causality, so the latter gives us no good reason to believe in the former.\footnote{See ref \cite{adlam2022roads} for a discussion of a similar error that crops up in discussions of locality in the interpretation of quantum mechanics}  

An alternative version of this defence involves the assertion that  scientific realism is in some way intrinsically linked to  causal explanation: for example, ref \cite{schmid2020unscrambling} states that `\emph{In our view, the most constructive way of defining ‘realities of Nature’ is as causal mechanisms acting on causal relata.}'  There is a Kantian element to this line of argument, since it seems to be proposing that causal concepts are an indispensable feature of any attempts we might make to formulate an idea of `reality.'  Indeed, Nagel has expressed a more explicitly Kantian version of this view, writing that causation `\emph{is an analytic consequence of what is commonly meant by `theoretical science' ... it is difficult to understand how it would be possible for modern theoretical science to surrender the general ideal expressed by the principle without becoming thereby transformed into something incomparably different from what that enterprise actually is}.'\cite{Nagel1961-NAGTSO-3} But history has shown that we should be cautious about assenting to these kinds of transcendental claims: after all, at one time Euclidean geometry\cite{friedman1992kant} and locality\cite{Einsteinlocal} were also both regarded as indispensable features of our thought, but science has since dispensed with both in certain regimes, without thereby becoming unintelligible. So   we should not be too quick to assume that causation is an essential feature of our attempts to describe reality. 

And finally it should be recognised that causal models for quantum mechanics retain the classical notion of `causation' only in a somewhat superficial way.  For if there exists something fundamentally causal underlying quantum mechanics,  it is not   really like classical strong causation at all -  for example, if the underlying fundamental causal structure is asymmetric, that asymmetry cannot be understood in interventionist terms, nor in perspectival terms, nor by appeal to the thermodynamic gradient, and therefore the causal relations involved must be inherently asymmetric in a way that we have apparently not seen in any other natural phenomena. So far from preserving a classical concept, this approach in fact introduces something radically new, and therefore it cannot really be justified  by appeal to conceptual conservatism.

 \section{Conclusion} 
 
 We have seen that the emerging research programme around the process matrix formalism helps clarify an important philosophical distinction between \emph{causal order} and \emph{strong causation}. Unlike strong causation, causal order can be defined without appeal to any notion of asymmetry, so it is possible for causal order to be present in fundamental physics even if it is perfectly time-symmetric, and indeed there is good evidence that   causal order is  instantiated by all physically possible singalling processes, or at least all signalling processes that are possible in the regimes we can currently access.

 This observation also suggests an account of the physical origins of strong causation: once we choose a direction, the bi-order associated with causal ordering becomes a directed acyclic graph, so  causal order can be combined with a perspectival or interventionist approach   in order to explain the appearance of strong causation at the macroscopic level. Without underlying causal order, the relationships between events would in general be too chaotic for any global macroscopic causal structure to appear. However, it must be emphasized that causal order is not the same as strong causation and it is important to keep the two notions separate to avoid confusion. Indeed, we would even be inclined to advocate that some word other than `causal/causation' should be used to describe what we have in this article referred to as `causal order,' since there are a number of important  disanalogies between causal order and our everyday concept of macroscopic causation, and this distinction is currently being obscured by the fact that the word `causal' is used throughout the literature to refer to both of these notions. 
 
 Finally, these results have important consequences for the way we should think about the foundations of physics. In particular, if we accept the account offered here for the origins of macroscopic strong causation, it follows that we should not expect to find natural causal accounts of the Bell correlations and other no-signalling quantum processes, since such processes are not subject to any ordering requirements and thus they do not  instantiate  causal order. This casts some doubt on the strategies pursued in various causal modelling approaches to quantum foundations, because it is unclear that we are justified in assuming that quantum mechanics has an underlying causal structure.
 
 We finish with the following question:   if there can be no causal account of the Bell correlations, how exactly should one think about the nature of  `nonlocality' in quantum physics? A detailed exploration of this topic must be left for future work, but roughly speaking we would expect that `locality' would emerge in a similar way to `strong causation.'  For  if we accept that all signalling processes must instantiate causal order, then since causal order  largely coincides with the lightcone structure of spacetime (indeed, we suggested in section \ref{science} that the lightcone structure might emerge from the causal structure),  it follows that once we choose a direction through the underlying bi-order, we will find that chains of  signalling processes will necessarily be associated with sequences of variables lying along a timelike or lightlike path in spacetime. Therefore chains of signalling processes will appear to be locally mediated from our macroscopic point of view, and yet this form of `locality' need  not be regarded as a fundamental feature of reality, but simply as an illusion which arises from the need for signalling processes to exhibit causal order.

 \section{Acknowledgements} 
 
 Thanks to Pascal Rodriguez Warnier for very helpful comments on a draft of this article.  This publication was made possible through the support of the ID 61466 grant from the John Templeton Foundation, as part of the “The Quantum Information Structure of Spacetime (QISS)” Project (qiss.fr). The opinions expressed in this publication are those of the author  and do not necessarily reflect the views of the John Templeton Foundation.

 \appendix
 
 \section{ Time Reversal \label{app}} 

Our description of time reversal has been quite schematic, so let us now be more precise. In fact the technical details of how to  implement an appropriate reversal operation are non-trivial, because this requires making a clean separation between the time-symmetric underlying physics and the asymmetries introduced by assumptions about the role of agents, and these features can be difficult to distinguish. Currently, it is known that at least some causal processes are time-reversal invariant in a sense made precise by Pienaar\cite{https://doi.org/10.48550/arxiv.1902.00129}.   Given an   IS description of a quantum process, Pienaar shows how to write down the corresponding OS process  using symmetric informationally complete (SIC) quantum instruments, and then proves that if the directed acyclic graph G associated with the process in the IS description has a certain structure (a `layered' structure) and the  process matrix W associated with the IS description is unbiased (inputting the maximally-mixed-state to the process produces the maximally-mixed state as output), then the process is  causally reversible, i.e.  the OS process is compatible with another IS descrption whose graph is obtained by reversing the direction of all the arrows in G. That is to say, for at least some causal processes it is indeed the case that we can reverse the IS description to obtain an alternative IS description which corresponds to the same underlying  OS  process, and moreover this reversed description will still be causal. It's unclear whether some result of this kind could be extended to all causal processes, but we do know that all physically possible processes can be obtained from `pure' processes by tracing out an ancilla system\cite{Araujo2017purification}, and `pure' processes are those which induce a unitary (and hence reversible) transformation from the past to the future whenever unitary transformations are also applied in the local laboratories; so it appears that causal processes can indeed be derived from time-reversal invariant underlying physics, and therefore  it would seem reasonable to conjecture that all causal processes should be reversible according to an appropriate formulation of operational time reversal, or at least can be extended using an ancilla to a process which is reversible in this way.

 That said,  we emphasise that our argument does not depend on this kind of conjecture.  Our aim here is to address the question of whether there can be causation in fundamental physics even if it is the case that, as many physicists currently believe, fundamental physics contains  no intrinsic temporal asymmetries. So the point we want to make is simply that the  notion of `causality' employed in the definition of `causal order' and `causal process' does not  \emph{need} an  underlying asymmetry:  processes may instantiate a well-defined causal order regardless of whether or not they can be reversed to yield another causally ordered process. Pienaar's result demonstrates that at least some causal processes can indeed be reversed in this way, thus verifying that the definition of `causal' does not depend in any crucial way on asymmetry.

  \bibliographystyle{unsrt}
 \bibliography{newlibrary12}{}

\begin{thebibliography}{10}

\bibitem{10.2307/4543833}
Bertrand Russell.
\newblock On the notion of cause.
\newblock {\em Proceedings of the Aristotelian Society}, 13:1--26, 1912.

\bibitem{10.2307/2215337}
Nancy Cartwright.
\newblock Causal laws and effective strategies.
\newblock {\em Noûs}, 13(4):419--437, 1979.

\bibitem{Woodward2007-WOOCWA}
James Woodward.
\newblock Causation with a human face.
\newblock In Huw Price and Richard Corry, editors, {\em Causation, Physics, and
  the Constitution of Reality: Russell's Republic Revisited}. Oxford University
  Press, 2007.

\bibitem{Field2003-FIECIA}
Hartry Field.
\newblock Causation in a physical world.
\newblock In Michael~J. Loux and Dean~W. Zimmerman, editors, {\em The Oxford
  Handbook of Metaphysics}, pages 435--460. Oxford University Press, 2003.

\bibitem{Norton2003-NORCAF}
John Norton.
\newblock Causation as folk science.
\newblock {\em Philosophers' Imprint}, 3:1--22, 2003.

\bibitem{Lewis1979-LEWCDA}
David Lewis.
\newblock Counterfactual dependence and time?s arrow.
\newblock {\em No\^us}, 13(4):455--476, 1979.

\bibitem{10.2307/20012433}
Phil Dowe.
\newblock Process causality and asymmetry.
\newblock {\em Erkenntnis (1975-)}, 37(2):179--196, 1992.

\bibitem{pearl2009causality}
Judea Pearl.
\newblock {\em Causality}.
\newblock Cambridge university press, 2009.

\bibitem{woodward2005making}
James Woodward.
\newblock {\em Making things happen: A theory of causal explanation}.
\newblock Oxford university press, 2005.

\bibitem{Price2005-PRICP}
Huw Price.
\newblock Causal perspectivalism.
\newblock In Huw Price and Richard Corry, editors, {\em Causation, Physics, and
  the Constitution of Reality: Russell's Republic Revisited}. Oxford University
  Press, 2005.

\bibitem{10.1086/678313}
James Woodward.
\newblock A functional account of causation; or, a defense of the legitimacy of
  causal thinking by reference to the only standard that matters—usefulness
  (as opposed to metaphysics or agreement with intuitive judgment).
\newblock {\em Philosophy of Science}, 81(5):691--713, 2014.

\bibitem{reversible}
Ämin Baumeler, Fabio Costa, Timothy Ralph, Stefan Wolf, and Magdalena Zych.
\newblock Reversible time travel with freedom of choice.
\newblock {\em Classical and Quantum Gravity}, 36, 11 2019.

\bibitem{Oreshkov2}
O.~{Oreshkov}, F.~{Costa}, and {\v C}.~{Brukner}.
\newblock {Quantum correlations with no causal order}.
\newblock {\em Nature Communications}, 3:1092, October 2012.

\bibitem{Oreshkov}
O.~{Oreshkov} and C.~{Giarmatzi}.
\newblock {Causal and causally separable processes}.
\newblock {\em New J. Phys}.

\bibitem{Ara_jo_2017}
Mateus Ara{\'{u} }jo, Adrien Feix, Miguel Navascu{\'{e}}s, and {\v{C}}aslav
  Brukner.
\newblock A purification postulate for quantum mechanics with indefinite causal
  order.
\newblock {\em Quantum}, 1:10, apr 2017.

\bibitem{rubinorozema}
Giulia Rubino, Lee Rozema, Adrien Feix, Mateus Araújo, Jonas Zeuner, Lorenzo
  Procopio, Časlav Brukner, and Philip Walther.
\newblock Experimental verification of an indefinite causal order.
\newblock {\em Science Advances}, 3, 08 2016.

\bibitem{Goswami_2018}
K.~Goswami, C.~Giarmatzi, M.~Kewming, F.~Costa, C.~Branciard, J.~Romero, and
  A.{\hspace{0.167em} }G. White.
\newblock Indefinite causal order in a quantum switch.
\newblock {\em Physical Review Letters}, 121(9), aug 2018.

\bibitem{articleoreshkov}
Ognyan Oreshkov and Christina Giarmatzi.
\newblock Causal and causally separable processes.
\newblock {\em New Journal of Physics}, 18, 09 2016.

\bibitem{Branciard_2015}
Cyril Branciard, Mateus Ara{\'{u} }jo, Adrien Feix, Fabio Costa, and
  {\v{C}}aslav Brukner.
\newblock The simplest causal inequalities and their violation.
\newblock {\em New Journal of Physics}, 18(1):013008, dec 2015.

\bibitem{renato}
V.~Vilasini and Renato Renner.
\newblock Embedding cyclic causal structures in acyclic spacetimes: no-go
  results for process matrices, 2022.

\bibitem{Lewis1986-LEWPTC}
David Lewis.
\newblock {Postscripts to `Causation'}.
\newblock In David Lewis, editor, {\em Philosophical Papers Vol. Ii}. Oxford
  University Press, 1986.

\bibitem{sep-causation-counterfactual}
Peter Menzies and Helen Beebee.
\newblock {Counterfactual Theories of Causation}.
\newblock In Edward~N. Zalta, editor, {\em The {Stanford} Encyclopedia of
  Philosophy}. Metaphysics Research Lab, Stanford University, {W}inter 2020
  edition, 2020.

\bibitem{sep-qm-collapse}
Giancarlo Ghirardi.
\newblock Collapse theories.
\newblock In Edward~N. Zalta, editor, {\em The Stanford Encyclopedia of
  Philosophy}. Metaphysics Research Lab, Stanford University, spring 2016
  edition, 2016.

\bibitem{https://doi.org/10.48550/arxiv.0906.2718}
David Wallace.
\newblock {A formal proof of the Born rule from decision-theoretic
  assumptions}, 2009.

\bibitem{Wallacebook}
D.~Wallace.
\newblock {\em The Emergent Multiverse: Quantum Theory according to the Everett
  Interpretation}.
\newblock OUP Oxford, 2012.

\bibitem{spirtes2000causation}
P.~Spirtes, R.~Scheines, and C.~Glymour.
\newblock {\em Causation, Prediction, and Search}.
\newblock Adaptive computation and machine learning. MIT Press, 2000.

\bibitem{https://doi.org/10.48550/arxiv.1309.6836}
Antti Hyttinen, Patrik~O. Hoyer, Frederick Eberhardt, and Matti Jarvisalo.
\newblock Discovering cyclic causal models with latent variables: A general
  sat-based procedure, 2013.

\bibitem{Neal_2000}
R.~M. Neal.
\newblock On deducing conditional independence from d-separation in causal
  graphs with feedback (research note).
\newblock {\em Journal of Artificial Intelligence Research}, 12:87--91, mar
  2000.

\bibitem{SpekkensWood}
C.~J. {Wood} and R.~W. {Spekkens}.
\newblock {The lesson of causal discovery algorithms for quantum correlations:
  causal explanations of Bell-inequality violations require fine-tuning}.
\newblock {\em New Journal of Physics}, 17(3):033002, March 2015.

\bibitem{Chaves}
R.~{Chaves}, C.~{Majenz}, and D.~{Gross}.
\newblock {Information-theoretic implications of quantum causal structures}.
\newblock {\em Nature Communications}, 6:5766, January 2015.

\bibitem{Miklin_2017}
Nikolai Miklin, Alastair~A Abbott, Cyril Branciard, Rafael Chaves, and
  Costantino Budroni.
\newblock The entropic approach to causal correlations.
\newblock {\em New Journal of Physics}, 19(11):113041, nov 2017.

\bibitem{Pienaar_2017}
Jacques Pienaar.
\newblock Which causal structures might support a quantum{\textendash}classical
  gap?
\newblock {\em New Journal of Physics}, 19(4):043021, apr 2017.

\bibitem{https://doi.org/10.48550/arxiv.1906.10726}
Jonathan Barrett, Robin Lorenz, and Ognyan Oreshkov.
\newblock Quantum causal models, 2019.

\bibitem{Chiribella_2013}
Giulio Chiribella, Giacomo~Mauro D'Ariano, Paolo Perinotti, and Benoit Valiron.
\newblock Quantum computations without definite causal structure.
\newblock {\em Physical Review A}, 88(2), aug 2013.

\bibitem{MacLean_2017}
Jean-Philippe~W. MacLean, Katja Ried, Robert~W. Spekkens, and Kevin~J. Resch.
\newblock Quantum-coherent mixtures of causal relations.
\newblock {\em Nature Communications}, 8(1), may 2017.

\bibitem{Gu_rin_2018}
Philippe~Allard Gu{\'{e} }rin and {\v{C}}aslav Brukner.
\newblock Observer-dependent locality of quantum events.
\newblock {\em New Journal of Physics}, 20(10):103031, oct 2018.

\bibitem{Paunkovi__2020}
Nikola Paunkovi{\'{c} } and Marko Vojinovi{\'{c}}.
\newblock Causal orders, quantum circuits and spacetime: distinguishing between
  definite and superposed causal orders.
\newblock {\em Quantum}, 4:275, may 2020.

\bibitem{doi:10.1098/rspa.2018.0903}
Giulio Chiribella and Hlér Kristjánsson.
\newblock Quantum shannon theory with superpositions of trajectories.
\newblock {\em Proceedings of the Royal Society A: Mathematical, Physical and
  Engineering Sciences}, 475(2225):20180903, 2019.

\bibitem{Costa_2022}
Fabio Costa.
\newblock A no-go theorem for superpositions of causal orders.
\newblock {\em Quantum}, 6:663, mar 2022.

\bibitem{10.3389/fphy.2020.525333}
Aleksandra Dimić, Marko Milivojević, Dragoljub Gočanin, Natália~S. Móller,
  and Časlav Brukner.
\newblock Simulating indefinite causal order with rindler observers.
\newblock {\em Frontiers in Physics}, 8, 2020.

\bibitem{10.2307/25592014}
Mathias Frisch.
\newblock 'the most sacred tenet'? causal reasoning in physics.
\newblock {\em The British Journal for the Philosophy of Science},
  60(3):459--474, 2009.

\bibitem{kutach2013causation}
D.~Kutach.
\newblock {\em Causation and Its Basis in Fundamental Physics}.
\newblock Oxford Studies in Philosophy of Science. OUP USA, 2013.

\bibitem{https://doi.org/10.48550/arxiv.1902.00129}
Jacques Pienaar.
\newblock A time-reversible quantum causal model, 2019.

\bibitem{Pienaar_2020}
Jacques Pienaar.
\newblock Quantum causal models via quantum bayesianism.
\newblock {\em Physical Review A}, 101(1), jan 2020.

\bibitem{https://doi.org/10.48550/arxiv.2206.02945}
Ken Wharton and Raylor Liu.
\newblock Entanglement and the path integral, 2022.

\bibitem{doi:10.1093/bjps/axp029}
Mathias Frisch.
\newblock {‘The Most Sacred Tenet’? Causal Reasoning in Physics}.
\newblock {\em The British Journal for the Philosophy of Science},
  60(3):459--474, 2009.

\bibitem{PhysRev.104.1760}
John~S. Toll.
\newblock Causality and the dispersion relation: Logical foundations.
\newblock {\em Phys. Rev.}, 104:1760--1770, Dec 1956.

\bibitem{1972causality}
{\em Causality and Dispersion Relations}.
\newblock ISSN. Elsevier Science, 1972.

\bibitem{10.2307/687169}
Graham Nerlich.
\newblock Special relativity is not based on causality.
\newblock {\em The British Journal for the Philosophy of Science},
  33(4):361--388, 1982.

\bibitem{doi:10.1063/1.1704140}
E.~C. Zeeman.
\newblock Causality implies the lorentz group.
\newblock {\em Journal of Mathematical Physics}, 5(4):490--493, 1964.

\bibitem{malament1977causal}
David Malament.
\newblock Causal theories of time and the conventionality of simultaneity.
\newblock {\em No{\^u}s}, pages 293--300, 1977.

\bibitem{Winnie1977-WINTCT}
John~A. Winnie.
\newblock The causal theory of space-time.
\newblock In John Earman, Clark Glymour, and John Stachel, editors, {\em
  Foundations of Space-Time Theories}. University of Minnesota Press, 1977.

\bibitem{Sorkingeometry}
R.~{ Sorkin}.
\newblock {Geometry from order: Causal sets}.
\newblock {\em Einstein Online}, 2006.

\bibitem{pittphilsci20539}
Emily Adlam.
\newblock {The Inaccessibility of the Past is not Statistical}, May 2022.

\bibitem{dibiagio2020quantum}
Andrea~Di Biagio, Pietro Donà, and Carlo Rovelli.
\newblock Quantum information and the arrow of time, 2020.

\bibitem{Hardy_2009}
Lucien Hardy.
\newblock Quantum gravity computers: On the theory of computation with
  indefinite causal structure.
\newblock In {\em The Western Ontario Series in Philosophy of Science}, pages
  379--401. Springer Netherlands, 2009.

\bibitem{universe7010012}
Jean-Pierre Luminet.
\newblock Closed timelike curves, singularities and causality: A survey from
  gödel to chronological protection.
\newblock {\em Universe}, 7(1), 2021.

\bibitem{10.2307/25592033}
Alyssa Ney.
\newblock Physical causation and difference-making.
\newblock {\em The British Journal for the Philosophy of Science},
  60(4):737--764, 2009.

\bibitem{schmid2020unscrambling}
David Schmid, John~H. Selby, and Robert~W. Spekkens.
\newblock Unscrambling the omelette of causation and inference: The framework
  of causal-inferential theories, 2020.

\bibitem{van1980scientific}
B.C. van Fraassen, Oxford~University Press, and P.P.B.C. Van~Fraassen.
\newblock {\em The Scientific Image}.
\newblock Clarendon Library of Logic and Philosophy. Clarendon Press, 1980.

\bibitem{Woodward2014}
James Woodward.
\newblock A functional account of causation; or, a defense of the legitimacy of
  causal thinking by reference to the only standard that matters—usefulness
  (as opposed to metaphysics or agreement with intuitive judgment).
\newblock {\em Philosophy of Science}, 81(5):691--713, 2014.

\bibitem{Lewis1973-LEWC}
David Lewis.
\newblock Causation.
\newblock {\em Journal of Philosophy}, 70(17):556--567, 1973.

\bibitem{adlam2021laws}
Emily Adlam.
\newblock {Laws of Nature as Constraints}.
\newblock {\em Foundations of Physics}, 52, 2022.

\bibitem{chen2021governing}
Eddy~Keming Chen and Sheldon Goldstein.
\newblock {Governing Without A Fundamental Direction of Time: Minimal
  Primitivism about Laws of Nature}, 2021.

\bibitem{Lewis1986-LEWOTP-3}
David Lewis.
\newblock {\em On the Plurality of Worlds}.
\newblock Wiley-Blackwell, 1986.

\bibitem{lewishumean}
David Lewis.
\newblock Humean supervenience debugged.
\newblock {\em Mind}, 103(412):473--490, 1994.

\bibitem{Armstrong1978-ARMATO}
David~Malet Armstrong.
\newblock {\em A Theory of Universals. Universals and Scientific Realism Volume
  Ii}.
\newblock Cambridge University Press, 1978.

\bibitem{10.2307/40230714}
Michael Tooley.
\newblock The nature of laws.
\newblock {\em Canadian Journal of Philosophy}, 7(4):667--698, 1977.

\bibitem{ismael2016physics}
J.T. Ismael.
\newblock {\em How Physics Makes Us Free}.
\newblock Oxford University Press, 2016.

\bibitem{Barrett_2021}
Jonathan Barrett, Robin Lorenz, and Ognyan Oreshkov.
\newblock Cyclic quantum causal models.
\newblock {\em Nature Communications}, 12(1), feb 2021.

\bibitem{2008PhRvA..77b2104M}
A.~{Montina}.
\newblock {Exponential complexity and ontological theories of quantum
  mechanics}.
\newblock {\em Physical Review A}, 77(2):022104, February 2008.

\bibitem{Adlamspooky}
Emily Adlam.
\newblock {Spooky Action at a Temporal Distance}.
\newblock {\em Entropy}, 20(1):41, 2018.

\bibitem{https://doi.org/10.48550/arxiv.2109.12128}
V.~Vilasini and Roger Colbeck.
\newblock A general framework for cyclic and fine-tuned causal models and their
  compatibility with space-time, 2021.

\bibitem{adlam2022roads}
Emily Adlam.
\newblock {Two Roads to Retrocausality}, 2022.

\bibitem{Nagel1961-NAGTSO-3}
Ernest Nagel.
\newblock {\em The Structure of Science: Problems in the Logic of Scientific
  Explanation}.
\newblock New York, NY, USA: Harcourt, Brace \& World, 1961.

\bibitem{friedman1992kant}
M.~Friedman.
\newblock {\em Kant and the Exact Sciences}.
\newblock Harvard University Press, 1992.

\bibitem{Einsteinlocal}
A.~Einstein.
\newblock Quanten-mechanik und wirklichkeit.
\newblock {\em Dialectica}, 2:320 -- 324, 05 2007.

\bibitem{sep-scientific-realism}
Anjan Chakravartty.
\newblock Scientific realism.
\newblock In Edward~N. Zalta, editor, {\em The Stanford Encyclopedia of
  Philosophy}. Metaphysics Research Lab, Stanford University, summer 2017
  edition, 2017.

\bibitem{Araujo2017purification}
Mateus Ara{\'{u}}jo, Adrien Feix, Miguel Navascu{\'{e}}s, and {\v{C}}aslav
  Brukner.
\newblock A purification postulate for quantum mechanics with indefinite causal
  order.
\newblock {\em {Quantum}}, 1:10, April 2017.

\end{thebibliography}

 \end{document}